\documentclass[12pt]{article}
\usepackage{epsfig}
\usepackage{hyperref}
\begin{document}

\title
{                   
\begin{flushright}{\normalsize HU-EP-10/37 \\
                               LU-ITP 2010/01 \\
                               BI-TP 2010/21  \\} 
\end  {flushright}\vspace{5mm}
Two-point functions of quenched lattice QCD in Numerical Stochastic Perturbation Theory.
\\
(II) The gluon propagator in Landau gauge}

\author
{F.~Di~Renzo  \\
\small
{Dipartimento di Fisica, Universit\`a di Parma and INFN, I-43100 Parma, Italy}
\and E.-M.~Ilgenfritz  \\
\small
{Fakult\"at f\"ur Physik, Universit\"at Bielefeld, D-33501 Bielefeld, Germany} \\
\small
{Institut f\"ur Physik, Humboldt-Universit\"at zu Berlin, D-12489 Berlin, Germany}
\and
H.~Perlt, A.~Schiller  \\
\small
{Institut f\"ur Theoretische Physik, Universit\"at Leipzig, D-04009 Leipzig, Germany}
\and
C.~Torrero  \\
\small
{Institut f\"ur Theoretische Physik, Universit\"at Regensburg, D-93040 Regensburg, Germany}
}

\maketitle

\begin{abstract}
This is the second of two papers devoted to the perturbative computation of the ghost and 
gluon propagators in $SU(3)$ Lattice Gauge Theory. Such a computation should enable a 
comparison with results from lattice simulations in order to reveal 
the genuinely non-perturbative content of the latter. The gluon propagator is computed 
by means of Numerical Stochastic Perturbation Theory: results range from two up to four 
loops, depending on the different lattice sizes. 
The non-logarithmic constants for one, two and three loops are extrapolated 
to the lattice spacing $a \to 0$ continuum and infinite volume $V \to \infty$ limits.
\end{abstract}

\section{Introduction}
In~\cite{DiRenzo:2009ni} a three-loop computation of the ghost propagator was presented.
As announced, we now report on a similar computation of the gluon propagator.
The full propagators of gluons and ghosts calculated on the lattice encode
information about the non-perturbative vacuum properties of QCD and of pure
Yang-Mills theory~\cite{Alkofer:2008bs}. This requires a non-perturbative
extension of the Landau gauge.

Although BRST invariance is essential for other non-perturbative approaches to non-Abelian
gauge theory, there are principal difficulties to reconcile it with present-day lattice gauge
fixing technology (as applied for calculating gauge-variant objects like gluon and ghost
propagators) in a way avoiding the Neuberger problem (see \cite{vonSmekal:2007ns} and references
therein). So far, there is no generally accepted way to deal with the Gribov ambiguity in lattice
simulations.

Other non-perturbative methods for calculations in the Landau gauge like
Schwinger-Dyson (SD) equations~\cite{Alkofer:2000wg} or the Functional Renormalization
Group (FRG)~\cite{Berges:2000ew,Pawlowski:2005xe} are not explicitly taking into account (and
seem not to be affected by) the complication compared to the standard Faddeev-Popov procedure
that the functional integration should be restricted to the Gribov horizon.
Zwanziger has argued~\cite{Zwanziger:2003cf} that the form of the SD equations
will not be affected. Rather supplementary conditions restricting solutions should account for it.
The so-called scaling solution~\cite{von Smekal:1997is,von Smekal:1997vx} has been an early
result of the SD approach obtained for power-like solutions. The power behavior has
been further specified in~\cite{Lerche:2002ep}
and confirmed by (time-independent) stochastic quantization~\cite{Zwanziger:2001kw}.

Concerning lattice simulations, Zwanziger proposed~\cite{Zwanziger:1993dh} to add nonlocal
terms to the action that should control the simulation to stay within the Gribov horizon~\footnote{In
this first attempt a singular ghost propagator $\propto 1/p^4$ has been obtained.}.
In the context of the Schwinger-Dyson approach this possibility has been analytically discussed first
in Ref.~\cite{Huber:2009tx}.
Standard Monte Carlo simulations without these refinements have failed to reproduce the theoretically
preferred far-infrared asymptotics of the scaling solution and have supported instead the
so-called decoupling solution (see Ref.~\cite{Aguilar:2008xm} and references therein).
This solution has later been shown to be possible in the SD
and FRG approaches with suitable boundary conditions~\cite{Fischer:2008uz},
but at the expense of a conflict with global BRST invariance.

Although we can assume that NSPT remains in the vicinity of the trivial vacuum, we have to understand
the present situation with respect to Monte Carlo results for gluon and ghost propagators.
Fortunately, the momentum range we are interested in and where we are going to compare with Monte Carlo
lattice results is not influenced by the Gribov ambiguity and the way of gauge fixing~\cite{Sternbeck:2005tk}.
{}From various studies it is known, however, that the intermediate momentum range of ${\cal O}(1 {\rm~GeV})$
is not less important from the point of view of confinement physics,
as seen for the gluon propagator~\cite{Langfeld:2001cz} and the quark propagator~\cite{Bowman:2008qd}
in Landau gauge on the lattice. From the FRG approach it is known that
for the onset of confinement at finite temperature the mid-momentum region of the propagators is
important~\cite{Braun:2007bx,Braun:2010cy}.
This is the region where violation of positivity~\cite{Langfeld:2001cz,Sternbeck:2006cg,Bowman:2007du}
invalidates a conventional, particle-like interpretation of the gluon propagator.
Specific non-perturbative configurations (center vortices) have been found to be 
essential~\cite{Langfeld:2001cz,Bowman:2008qd} to understand the behavior of the gluon and the quark propagator in
the intermediate momentum range.

In recent years, also the large-momentum behavior of the lattice gluon and ghost propagators has
attracted growing interest, in particular with respect to the coupling constant of the ghost-gluon
vertex~\cite{Sternbeck:2007br,Boucaud:2008gn,Sternbeck:2010xu,Blossier:2010ky}, which has the potential
to provide an independent precision measurement of $\alpha_s(M_Z)$ from these propagators~\cite{vonSmekal:2009ae}.
First estimates of the zero- and two-flavor values of
$\Lambda_{\overline{\rm MS}}$~\cite{Boucaud:2008gn,Sternbeck:2010xu,Blossier:2010ky} and a possible
dimension-two condensate~\cite{Boucaud:2008gn,Blossier:2010ky} are available already and look promising.
Therefore, a detailed knowledge of the propagators' lattice perturbative part would much foster these
efforts.

When our work begun in 2007~\cite{Ilgenfritz:2007qj,DiRenzo:2007qf}, 
the intention was to clarify the perturbative background, among other facts the type of convergence 
of the summed-up few-loop perturbative contributions to the propagators in various momentum ranges. 
In standard Lattice Perturbation Theory  (LPT) such calculations are very difficult beyond two-loop order. 
To overcome this obstacle, we have used the method of Numerical Stochastic Perturbation
Theory (NSPT)~\cite{NSPT}, which provides a stochastic, automatized framework 
for gauge-invariant and gauge-non-invariant calculations. 
 
Stochastic gauge fixing is built in, and a high-precision procedure has been devised to fix 
the gauge to Landau, at any given order. 
Thus, the propagators (for each lattice momentum) can be calculated in subsequent 
orders of perturbation theory. A limit is set only by storage limitations and machine precision.
There are no truncation errors.
For direct comparison with Monte Carlo results, we present the low-loop results summed up with inverse 
powers of the bare inverse coupling $\beta^n$. 
In this paper we also try, for the first time applied
to the gluon propagator, to improve the convergence by applying boosted perturbation theory.

The effectiveness of NSPT relies 
on the fact that the parametrization of the (leading and non-leading) 
logarithmic terms can follow largely in accordance with standard perturbation theory. 
The essential difficulty left to NSPT is the computation of the constant contributions, 
which are in general very difficult to achieve in diagrammatic LPT. 

Only the one-loop constant term was known since long~\cite{Kawai:1980ja} for the ghost and gluon propagator. 
Reproducing these results, which are obtained in the continuum and infinite-volume limits, 
was the first feasibility test for NSPT~\cite{Ilgenfritz:2007qj}. 

In general, at any order, NSPT results are obtained at finite lattice spacing and finite volume. 
A fitting procedure is needed to get the continuum ($a \to 0$) and infinite-volume ($V \to \infty$) 
limits. While the extraction of the first (continuum) limit relies on hypercubic-invariant Taylor 
series~\cite{Di Renzo:2006wd}, a careful extraction of the second (infinite-volume) limit requires the 
accounting of $pL$ contributions ($p$ being the momentum scale relevant to the computation and 
$L$ the finite extent of the lattice). In the first paper (abbreviated as I), we gave a quite 
comprehensive description of all this technology while applying it to the ghost propagator. 
In the present, second paper we are going to apply the method to an analysis of the gluon propagator.     

The paper is organized as follows. 
Sect.~\ref{sec:gluonprop-formulation} recalls the lattice definition of the gluon propagator,
together with specific features that the calculation in the framework of NSPT contains.
Sect.~\ref{sec:standard-LPT} contains the nomenclature of standard Lattice Perturbation Theory
where our results have to fit in. In Sect.~\ref{sec:implementation} we only briefly describe 
the implementation of NSPT. The interested reader will find more information of this kind in 
part I of this series of papers. However, we document the statistics for different lattice 
volumes and different orders of perturbation theory that has been collected by the Leipzig 
and Parma part of the collaboration. Here we also present the raw data before and after the 
extrapolation to the Langevin time-step $\epsilon \to 0$ limit.
In Sect.~\ref{sec:comparison} we compare the results of NSPT (up to four loops) with Monte Carlo 
results and try to improve the convergence by boosted perturbation theory. 
Sect.~\ref{sec:fitting} presents the fitting procedure and the final results for the 
leading and non-leading loop corrections.
In Sect.~\ref{sec:summary} we draw our conclusions and summarize our results.

\section{The lattice gluon propagator}
\label{sec:gluonprop-formulation}

The lattice gluon propagator $D^{ab}_{\mu\nu}(p(k))$ is the Fourier transform
of the gluon two-point function, {\it i.e.} the expectation value
\begin{equation}
  D^{ab}_{\mu\nu}(p(k)) = \left\langle \widetilde{A}^a_{\mu}(k)
  \widetilde{A}^b_{\nu}(-k) \right\rangle = \delta^{ab} D_{\mu\nu}(p(k)) \,,
\label{eq:D-definition}
\end{equation}
which is required to be color-diagonal
and symmetric in the Lorentz indices $\mu,\nu$.
For the definition of the lattice momenta $p_\mu (k_\mu)$, $k_\mu$ 
and $\hat p_\mu(k_\mu)$ 
to be used later we refer to (I-17)-(I-19).

Assuming reality of the color components of the vector potential 
and rotational invariance of the two-point function, the
continuum gluon propagator has the following general tensor 
structure~\footnote{Differently to the other chapters
$p$ denotes here directly the continuum Euclidean four-momentum.}
\begin{equation}
  D_{\mu\nu}(p) = \left( \delta_{\mu\nu} - \frac{p_{\mu}~p_{\nu}}{p^2}
  \right) D(p^2) + \frac{ p_{\mu}~ p_{\nu}}{p^2}
        \frac{F(p^2)}{p^2} \; ,
  \label{eq:decomposition}
\end{equation}
with $D(p^2)$ and $F(p^2)$ being the transverse and longitudinal propagator,
respectively. The longitudinal propagator $F(p^2)$ vanishes in the Landau gauge.

The lattice gluon propagator $D_{\mu\nu}(p(k))$ depends on the lattice 
four-momentum $p(k)$.
Due to the lower symmetry of the hypercubic group its general tensor structure can 
be expected to be more complicated than~(\ref{eq:decomposition}) that holds in the continuum.
Inspired by the continuum form~(\ref{eq:decomposition}) we consider as one strategy only the 
extraction of the following lattice scalars 
\begin{eqnarray*}
 \sum_{\mu,\nu} \hat{p}_\mu(k_\mu) \, D_{\mu\nu}(p(k)) \, \hat{p}_\nu(k_\nu)\,, \quad 
  \sum_\mu D_{\mu\mu}(p(k))
\end{eqnarray*} 
that should survive the continuum limit.
Note, however, that additional lattice scalars could be measured as well.
The first scalar vanishes exactly in lattice Landau gauge.
In this gauge the second scalar function, corresponding to the transverse part 
of the gluon propagator in the continuum limit, is denoted by
\begin{equation}
  D(p(k)) =\frac{1}{3} \sum^{4}_{\mu=1} D_{\mu\mu}(p(k)) \,.
  \label{eq:decomposition1}
\end{equation}
On the lattice, this function is influenced by the
the lower symmetry of the hypercubic group.

In NSPT the different loop orders $n$  (even orders in $\beta^{-1/2}$) 
at finite Langevin step size $\varepsilon$ are constructed directly from the
Fourier transformed perturbative gauge fields $\widetilde{A}^{a,(l)}_\mu(k)$
with (see (I-6) and (I-7))
\begin{equation}
  A_{x+\hat{\mu}/{2},\mu}= \sum_{l>0} \beta^{-l/2} A^{(l)}_{x+\hat{\mu}/{2},\mu} \,, \quad 
  A^{(l)}_{x+\hat{\mu}/{2},\mu}= \sum_a T^a A^{a,(l)}_{x+\hat{\mu}/{2},\mu} \,.
\end{equation}
what leads to
\begin{equation}
  \delta^{ab} D_{\mu\nu}^{(n)}(p(k)) = \left\langle \,
  \sum_{l=1}^{2n+1}
  \left[ \widetilde{A}^{a,(l)}_{\mu}(k) \,
  \widetilde{A}^{b,(2n+2-l)}_{\nu}(-k) \right] \,
  \right\rangle \,.
\label{eq:Dn}
\end{equation}

Note that already the tree-level contribution to the gluon propagator, 
$D_{\mu\nu}^{(0)}$, arises from quantum fluctuations of the gauge fields with $l=1$.
In addition, terms with non-integer $n=1/2,3/2,\dots$ in the previous equation (\ref{eq:Dn}) 
-- which do not correspond to loop contributions -- 
should vanish numerically after averaging over configurations.

Similar to the ghost propagator in paper I, we present the various orders of
the gluon dressing function (or ``form factor'') in two forms:
\begin{equation}
  J^{G,(n)}(p) = p^2 \; D^{(n)}(p(k)) \,, \quad
  \hat J^{G,(n)}(p) = \hat p^2 \; D^{(n)}(p(k)) \, .
\label{eq:dressingfunction}
\end{equation}
Contrary to the ghost propagator
the gluon dressing function can be calculated at the same time for all possible four-momenta
given  by four integers -- the four-momentum tuples $(k_1,k_2,k_3,k_4)$.
This makes a calculation of the gluon propagator significantly cheaper. 
The tree-level result for the dressing function, $\hat J^{G,(0)}(p(k))=1$
in the limit $\varepsilon \to 0$ for all sets of tuples
is non-trivial and is obtained as the result of averaging.

\section{The propagator and standard Lattice Perturbation Theory}
\label{sec:standard-LPT}

As discussed in paper I -- we relate infrared singularities
encountered in our finite volume NSPT calculation to powers of logarithms
of the external momentum obtained in the infinite volume limit.
Therefore, we need the anomalous dimension of the gluon field $A_\mu$, the
$\beta$-function and the relation between lattice bare and renormalized coupling. 
The procedure is outlined in detail in Section 4 of I. Here, we only repeat the
essential equations and quote the final numbers.
To avoid a possible mismatch of equations we add an index $G$ for the gluon propagator case.

In the RI'-MOM scheme, the renormalized gluon dressing function $J^{G,\rm RI'}$ is
defined as
\begin{equation}
  J^{G,\rm RI'}( p, \mu , \alpha_{\rm RI'}) =
  \frac{J^G(a, p, \alpha_{\rm RI'})}{Z^G(a,\mu,\alpha_{\rm RI'})} \, ,
  \label{eq:renorm}
\end{equation}
with the standard condition
\begin{equation}
  J^{G,\rm RI'}( p, \mu , \alpha_{\rm RI'})|_{p^2=\mu^2} = 1 \, .
  \label{eq:RIcondition}
\end{equation}
The gluon dressing function $J^G(a,p,\alpha_{\rm RI'})$ is the gluon
wave function renormalization constant
$Z^G(a,\mu,\alpha_{\rm RI'})$ at $\mu^2=p^2$.

The expansions of $Z^G$, $J^G$ and $J^{G,\rm RI'}$ in terms of the
renormalized coupling
$\alpha(a\mu)= (g(a\mu))^2/(16\pi^2)$ are completely analogous to (I-39)-(I-41).
The gluon wave function renormalization as an expansion in the lattice coupling  
$ \alpha_0={N_c}/{(8 \pi^2 \beta)} $ is represented as
\begin{equation}
  Z^G(a,\mu,\alpha_0)=1+ \sum_{i>0} \alpha_0^i \, \sum_{j=0}^i\,
  z^G_{i,j}\,\left(\log(a\mu)\right)^j\,.
  \label{ZAbare}
\end{equation}
This is the expansion we can measure in NSPT.

Again we restrict ourselves to three-loop expressions for the Landau gauge in the quenched approximation. 
The coefficients in front of the logarithms are partly known from calculations of the gluon wave functions 
and the beta function in the continuum and given as follows
(compare {\em e.g.} \cite{Gracey:2003yr})
\begin{eqnarray}
  &&
  z^{G,\rm RI'}_{1,1}= - \frac{13}{3} \, N_c \,,
  \\
  &&
  z^{G,\rm RI'}_{2,2}= - \frac{13}{2} \, N_c^2 \,, \quad
  z^{G,\rm RI'}_{2,1}= - \frac{3727 }{108}\, N_c^2 +
  3 \, z^{G,\rm RI'}_{1,0} \,N_c  \,,
  \\
  &&
  z^{G,\rm RI'}_{3,3}= - \frac{403}{18} \, N_c^3 \,, \quad
  z^{G,\rm RI'}_{3,2}= - \frac{5495 }{36} \, N_c^3
  + \frac{31}{2} \, z^{G,\rm RI'}_{1,0}\, N_c^2
  \,,
  \\
  &&
  z^{G,\rm RI'}_{3,1}=- \frac{2127823}{3888} \, N_c^3
  + \frac{361}{8} \, \zeta [3] \, N_c^3
  - \frac{1279}{108} \, z^{G,\rm RI'}_{1,0}\, N_c^2
  + \frac{31}{3} \, z^{G,\rm RI'}_{2,0}\, N_c\,.
  \nonumber
\end{eqnarray}
The finite one-loop constant $z^{G,\rm RI'}_{1,0}$ is known from the
gluon self energy in standard  infinite volume LPT~\cite{Kawai:1980ja} with the value
\begin{equation}
  z^{G,\rm RI'}_{1,0}=60.3673076 \, .
\end{equation}
Again the constants $z^{G,\rm RI'}_{2,0}$ and $z^{G,\rm RI'}_{3,0}$ were not known so far.
The form of the coefficients $z^G_{i,j}$ of $Z^G(a,\mu,\alpha_0)$ in (\ref{ZAbare}) up to three loops
can be read off directly from equations (I-51)-(I-53) replacing there all $z_{i,j}$ by $z^G_{i,j}$.

As a result, we present the gluon dressing function 
as function of the inverse lattice coupling $\beta$ to that order
\begin{eqnarray}
  J^{G,\rm 3-loop}(a,p, \beta) =
  1+\sum_{i=1}^{3} \frac{1}{\beta^{i}} J^{G,(i)}(a,p) \,, \quad
  J^{G,(i)}(a,p)=\sum_{j=0}^i\, J^G_{i,j}\,\left(\log (ap)^2 \right)^j
  \label{Zgluonbeta3loop}
  \,,
\end{eqnarray}
with the one-loop coefficients
\begin{eqnarray}
  \label{JG1loop}
  J^G_{1,1}&=&-0.24697038\,,
  \nonumber
  \\
  J^G_{1,0}&=&0.03799544  \,z^{G,\rm RI'}_{1,0}= 2.29368 \, ,
\end{eqnarray}
the two-loop coefficients
\begin{eqnarray}
  \label{JG2loop}
  J^G_{2,2}&=&0.08210781 \,,
  \nonumber \\
  J^G_{2,1}&=&-0.917978574 - 0.00938375 \, z^{G,\rm RI'}_{1,0}=-1.484450\,,
  \\
  J^G_{2,0}&=&0.10673710 \, z^{G,\rm RI'}_{1,0} + 0.00144365  \, z^{G,\rm RI'}_{2,0}
  =6.443431 + 0.00144365 \, z^{G,\rm RI'}_{2,0} \,,
  \nonumber
\end{eqnarray}
and the three-loop coefficients
\begin{eqnarray}
  \label{JG3loop}
  J^G_{3,3}&=&-0.02963736\,,
  \nonumber \\
  J^G_{3,2}&=&0.62856608 + 0.00311972\, z^{G,\rm RI'}_{1,0}=0.81689534\,,
  \nonumber \\
  J^G_{3,1}&=&-4.06861256 - 0.06123991\, z^{G,\rm RI'}_{1,0}- 0.00035654\, z^{G,\rm RI'}_{2,0}
  \nonumber \\
  &=&-7.7655009 - 0.00035654\, z^{G,\rm RI'}_{2,0}
  =-6.174164 - 0.246970 J^G_{2,0} \,,
  \\
  J^G_{3,0}&=&0.375990\, z^{G,\rm RI'}_{1,0}+0.00811105\, z^{G,\rm RI'}_{2,0}
  +0.0000548523\, z^{G,\rm RI'}_{3,0}
  \nonumber \\
  &=&22.697532 + 0.00811105\, z^{G,\rm RI'}_{2,0} +0.0000548523\, z^{G,\rm RI'}_{3,0}\,.
  \nonumber
\end{eqnarray}
Let us repeat it once more: the leading logarithmic coefficients for a given order can be exclusively
taken from continuum perturbative calculations.
The non-leading log coefficients are influenced, however, by the finite lattice constants 
from corresponding lower loop orders.

\section{Results of the NSPT calculations} 
\label{sec:implementation}

\subsection{Statistics}

To obtain infinite volume perturbative loop results at vanishing lattice spacing, we have to study again
the limit $\varepsilon \to 0$ and different lattice sizes $N$. 
We have used $N=6, 8, 10, 12$ and $N=16, 20, 32$ and studied the maximal loop order for the propagator 
$n_{\max}=4$ and $n_{\max}=2$, respectively.
The accumulated statistics for the different $\varepsilon$'s and lattice sizes 
are collected in Tables~\ref{tab:statistics} and~\ref{tab:statistics2}.
\begin{table}[!htb]
  \begin{center}
    \begin{tabular}{|c|c|c|c|c|c|}
\hline
  $\varepsilon$ & $N=6$ & $N=8$& $N=10$& $N=12$& $N=16$\\
\hline
    0.01        &  1500 & 750  & 3000   &  1000  &  3000\\  
    0.02        &  1000 & 750  & 2000   &  1000  &  2000\\  
    0.03        &  1000 & 750  & 2000   &  1000  &  2000\\  
    0.05        &  1000 & 750  & 2000   &  1000  &  2000\\  
    0.07        &  1000 & 750  & 2000   &  1000  &  2000\\  
\hline
    \end{tabular}
  \end{center}
  \caption{Number of gluon propagator measurements up to four loops ($N \le 12$) and up to one loop
           ($N=16$) using the Leipzig NSPT code.}
  \label{tab:statistics}
\end{table}
\begin{table}[!htb]
\begin{center}
\begin{tabular}{|c|c|c|c|}
\hline
  $\varepsilon$ & $N=16$  & $N=20$  &$N=32$  \\
\hline
    0.010       & 7436    & 5965 &810\\
    0.015       & 3053    & 3896 &\\
    0.020       & 4725    & 3015 &715\\
    0.040       & 2827    & 2633 &835\\
\hline
\end{tabular}
\end{center}
\caption{Number of tree-level, one- and two-loop gluon propagator measurements  at lattices sizes
         $N=16, 20, 32$ using the Parma NSPT code.}
\label{tab:statistics2}
\end{table}

The Landau gauge  was defined by the condition (I-14).
In the gluon propagator case  we have used in the gauge fixing condition (I-20)  
$l_{\max}=10$ orders of the perturbative gauge fields to obtain 
the propagator up to four loops in the case of the smaller volumes,
and used $l_{\max}=6$ for the bigger volumes with $N=16, 20, 32$.

\subsection{Raw data and check of vanishing contributions}

\begin{figure}[!htb]
  \begin{tabular}{cc}
     \includegraphics[scale=0.64,clip=true]{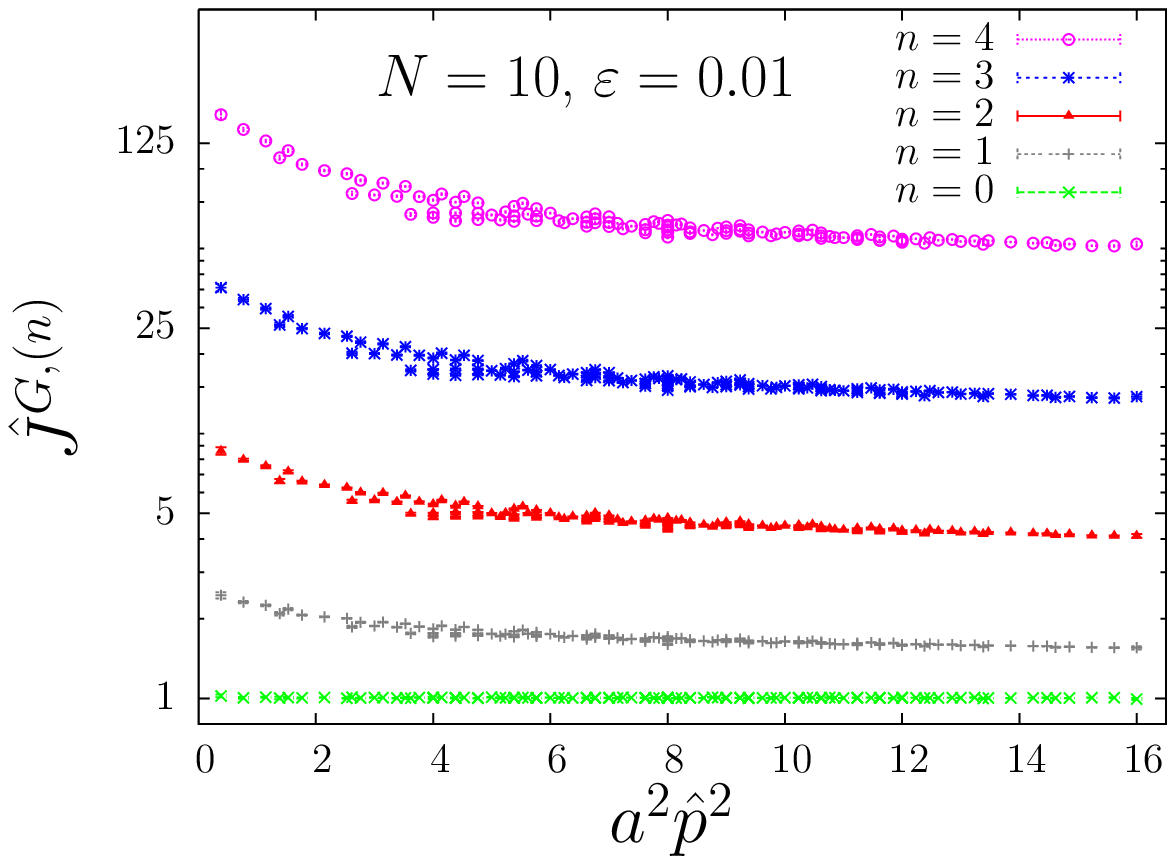}
     &
     \includegraphics[scale=0.64,clip=true]{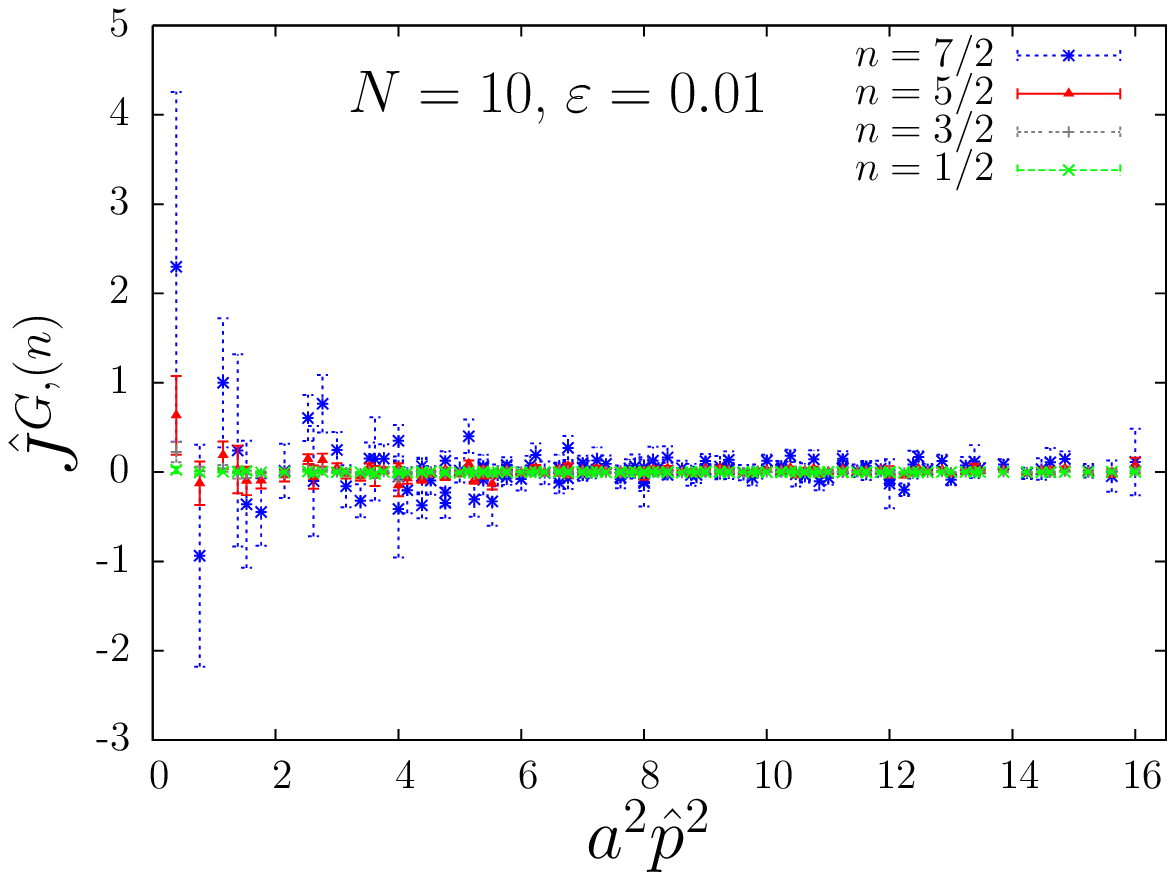}
  \end{tabular}
  \vspace{-2mm}
  \caption{Raw data $\hat J^{G,(n)}$  vs. $a^2\hat p^2$
           at $N=10$ and $\varepsilon=0.01$.
           Left: Increasing loop contributions from below. Right: Vanishing non-loop contributions.}
  \label{fig:raw}
\end{figure}

In Fig.~\ref{fig:raw}
we present as an example the measured dressing function $\hat J^{G,(n)}$
as function of $a^2\hat p^2$  for $n=0,1,2,3,4$ and $n=1/2,3/2,5/2,7/2$, respectively,
for a lattice size $10^4$ and the Langevin step $\varepsilon=0.01$.
Note that due to the hypercubic symmetry different realizations of the four-momentum tuples $k$ 
can lead to the same $a^2\hat p^2$ forming different branches.

In the left part of Fig.~\ref{fig:raw} the different contributions are shown starting from tree-level 
up to four loops.
The higher loop contributions are of the same sign leading to a monotonic increase of the perturbative
dressing function including higher and higher loop contributions.
The right part of Fig.~\ref{fig:raw} nicely demonstrates the vanishing of the 
non-loop contributions.
The size of the ``approximate zeros'' has to be compared with the corresponding 
sizes of the loop contributions of the ``adjacent'' integer loop orders. 

\begin{figure}[!htb]
  \begin{center}
    \includegraphics[scale=0.64,clip=true]{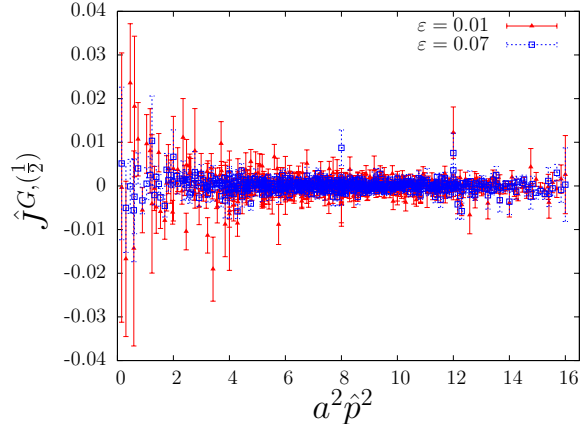}
   \end{center}
   \caption{$\hat J^{G,(1/2)}$ vs. $a^2\hat p^2$ at
            $N=16$ and $\varepsilon=0.01$ and $0.07$.}
  \label{fig:vanishing}
\end{figure}

Fig.~\ref{fig:vanishing}
demonstrates for a chosen $n=\frac{1}{2}$ that the corresponding 
averages equal to zero are indeed reached at {\sl finite} Langevin step size $\varepsilon$.
Typically, the errors increase with decreasing step size $\varepsilon$ and $a^2 \hat p^2$.

\begin{figure}[!htb]
  \begin{tabular}{cc}
    \includegraphics[scale=0.64,clip=true]{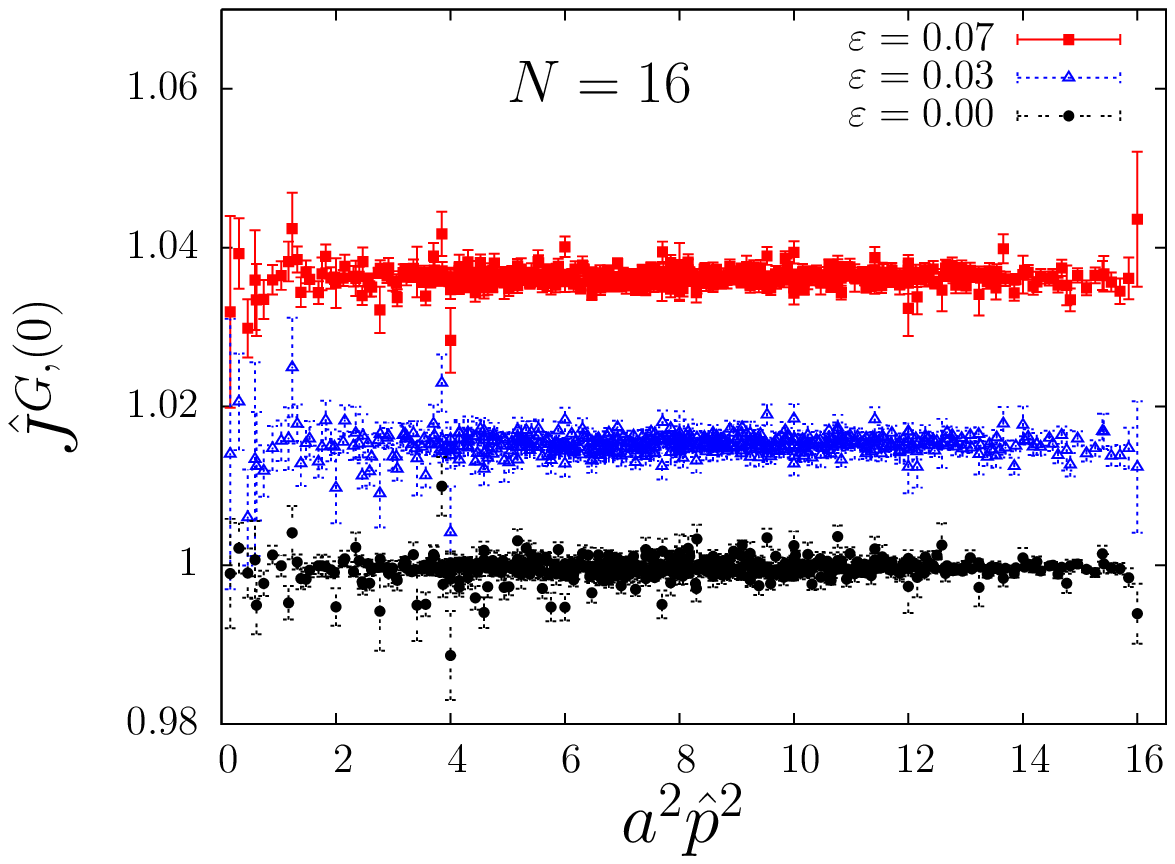}
    &
    \includegraphics[scale=0.64,clip=true]{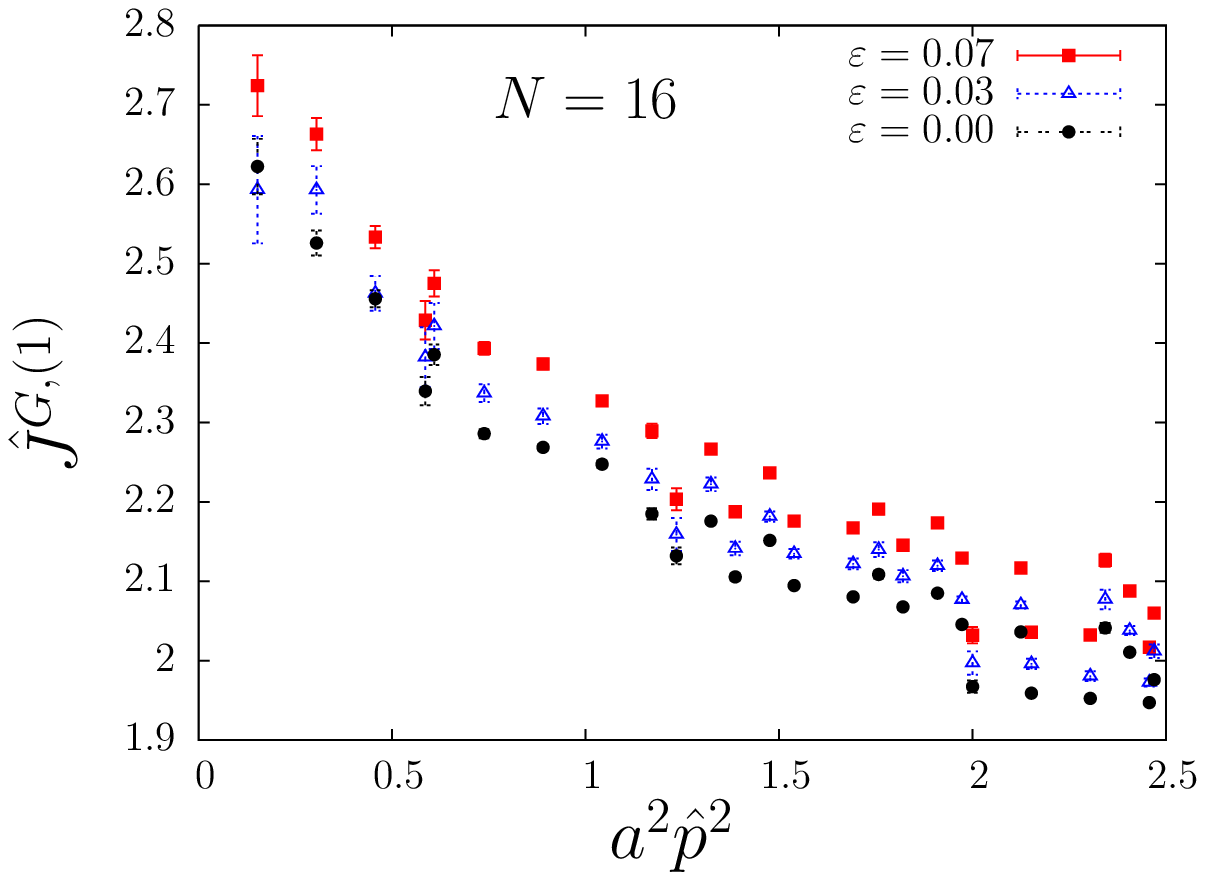}
  \end{tabular}
  \caption{The $\varepsilon \to 0$ extrapolation of the tree level $\hat J^{G,(0)}$ (left) 
           and one-loop $\hat J^{G,(1)}$ (right) dressing function vs. $a^2\hat p^2$ at  $N=16$.
           Note that the one-loop result is shown in a smaller $a^2\hat p^2$ window. 
           The $\varepsilon \to 0$ extrapolation is reached from above.}
  \label{fig:tree-extrapolation}
\end{figure}

In Fig.~\ref{fig:tree-extrapolation} 
we present the result of the linear extrapolation  with $\varepsilon \to 0$ 
for the tree-level $\hat{J}^{G,(0)}$  and one-loop $\hat{J}^{G,(1)}$ gluon dressing functions
on a $16^4$ lattice, together with 
the data of all inequivalent four-momentum tuples for $\varepsilon=0.07$ and $\varepsilon=0.03$, 
as a function of $a^2\hat{p}^2$. Note to 
what precision the expected result ``One'' for $\hat{J}^{G,(0)}$ is reproduced.

\begin{figure}[!htb]
  \begin{center}
    \begin{tabular}{cc}
       \includegraphics[scale=0.65,clip=true] {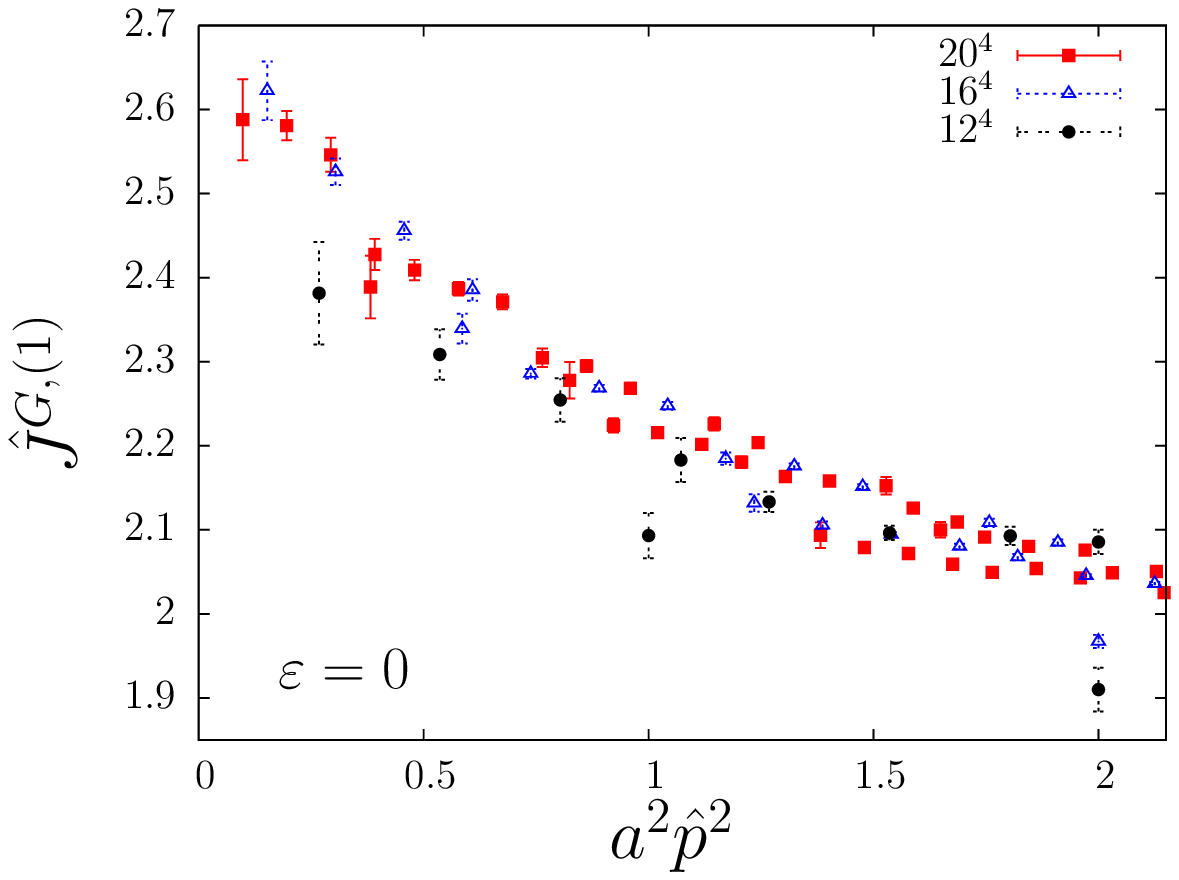}
       &
       \includegraphics[scale=0.65,clip=true] {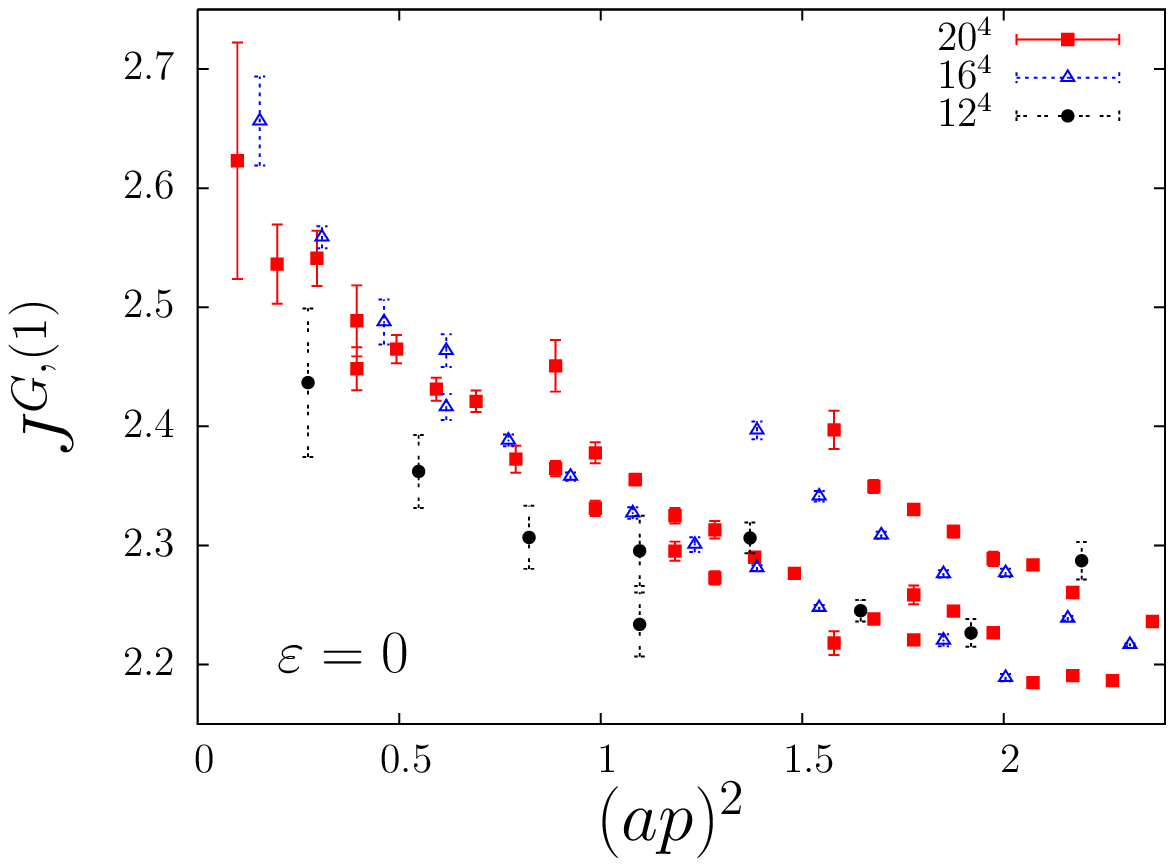}
    \end{tabular}
  \end{center}
  \caption{The one-loop dressing functions $\hat J^{G,(1)}$ (left) and $J^{G,(1)}$ (right)
           at selected different volumes vs. small momentum squared   $a^2 \hat p^2$ and $(ap)^2$.}
   \label{fig:vol-1loop}
\end{figure}

\begin{figure}[!htb]
  \begin{center}
    \begin{tabular}{cc}
      \includegraphics[scale=0.65,clip=true]{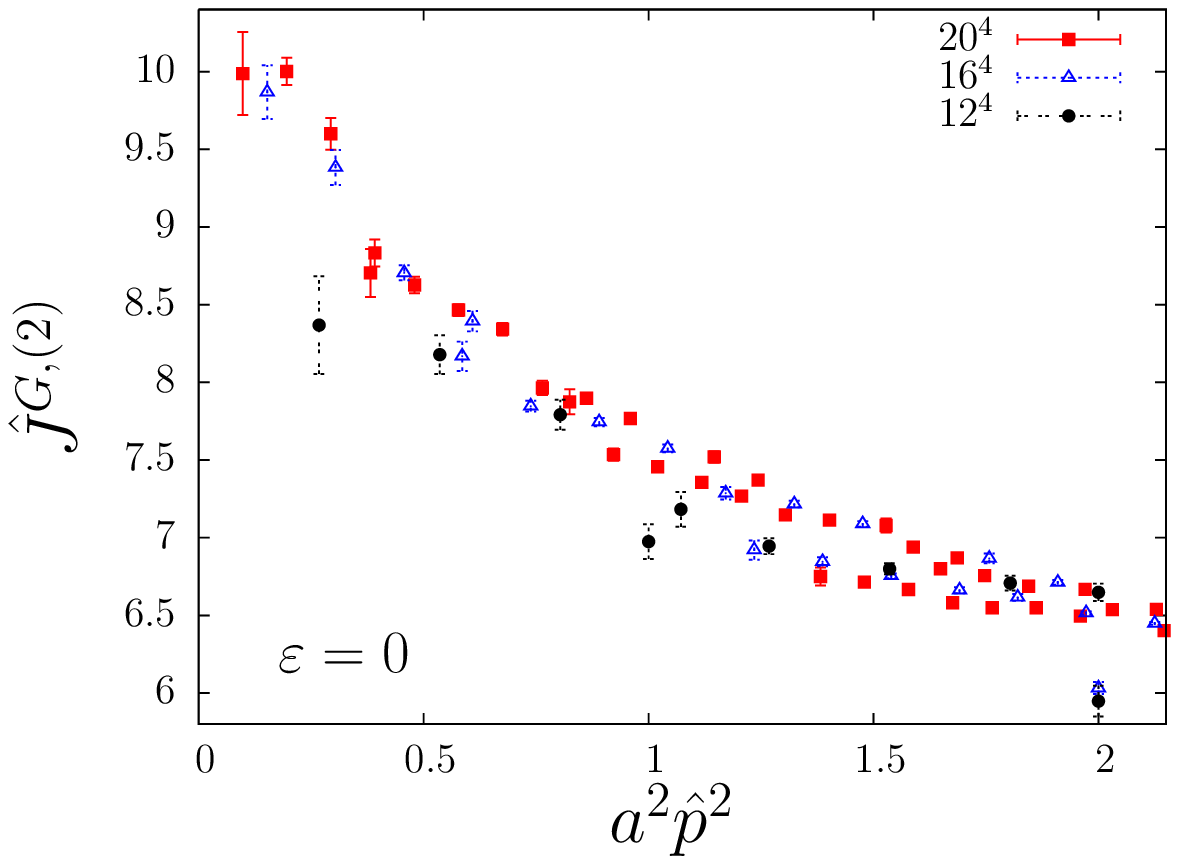}
      &
      \includegraphics[scale=0.65,clip=true]{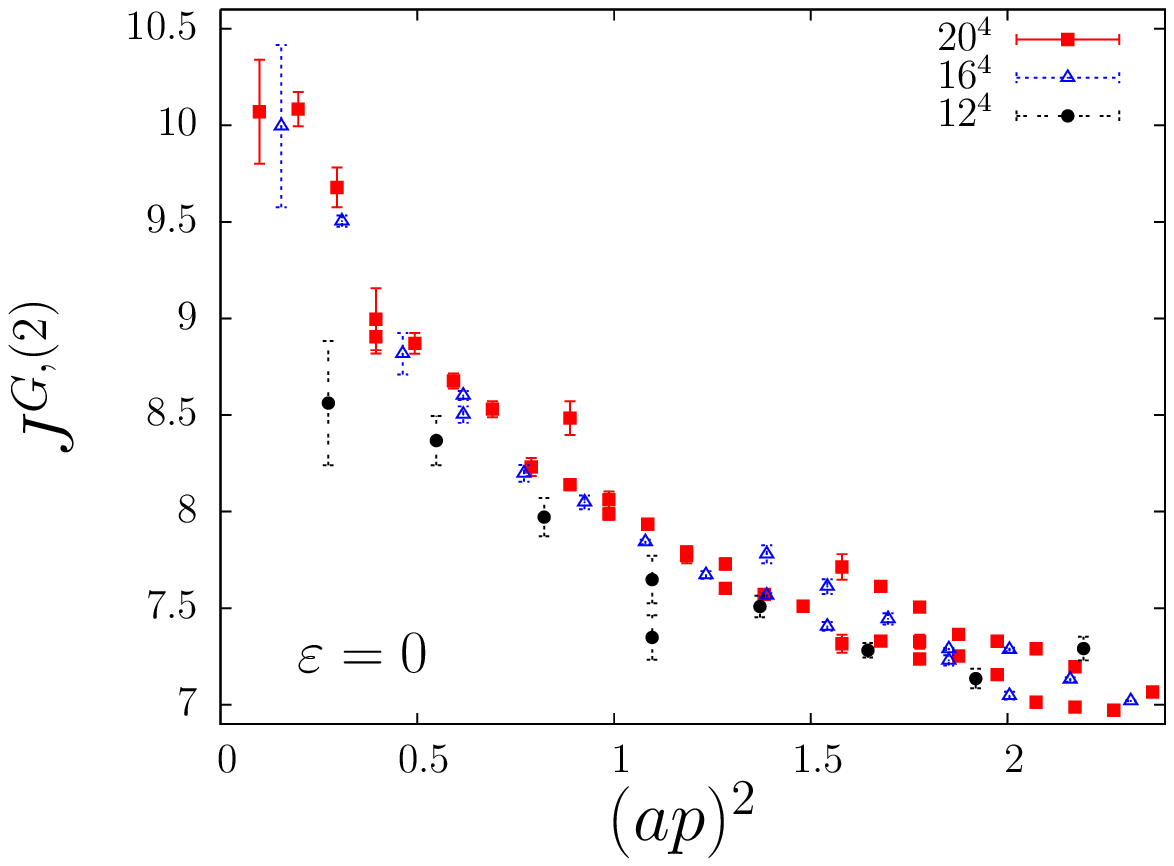}
    \end{tabular}
  \end{center}
  \caption{Same as in Fig.~\ref{fig:vol-1loop} for the two-loop dressing functions
           $\hat J^{G,(2)}$  and $J^{G,(2)}$.}
  \label{fig:vol-2loop}
\end{figure}

Figs.~\ref{fig:vol-1loop} and~\ref{fig:vol-2loop} 
show how  the one-loop and two-loop dressing functions
$\hat{J}^{G,(n)}$ and $J^{G,(n)}$ depend on $a^2 \hat p^2$ and $(ap)^2$, respectively. 
The different branches for the inequivalent off-diagonal four-momentum tuples in use are
clearly seen, which are strongly volume dependent.

\section{The perturbative gluon propagator summed to four loops}
\label{sec:comparison}

\subsection{Naive summation}

Having obtained the different loop contributions to the dressing function in Landau gauge, 
we can sum up those contributions for a given inverse gauge coupling $\beta$ to get an estimate 
of the perturbative gluon propagator.

We calculate the perturbative gluon dressing function at a given lattice volume
summed up to loop order $n_{\max}$ for a given lattice coupling $\beta$ as follows:
\begin{equation}
  \hat J^{G}(n_{\max})=  \sum_{n=0}^{n_{\max}} \frac{1}{\beta^n}
    \, \hat J^{G,(n)}
\end{equation}
and similar for $J^{G}(n_{\max})$.
In contrast to the ghost propagator, here also the tree level ``One'' is calculated numerically 
(not shown in the Figures below)
and we take the extracted numbers being near to the exact number one.

\begin{figure}[!htb]
  \begin{center}
    \begin{tabular}{cc}
       \includegraphics[scale=0.64,clip=true]{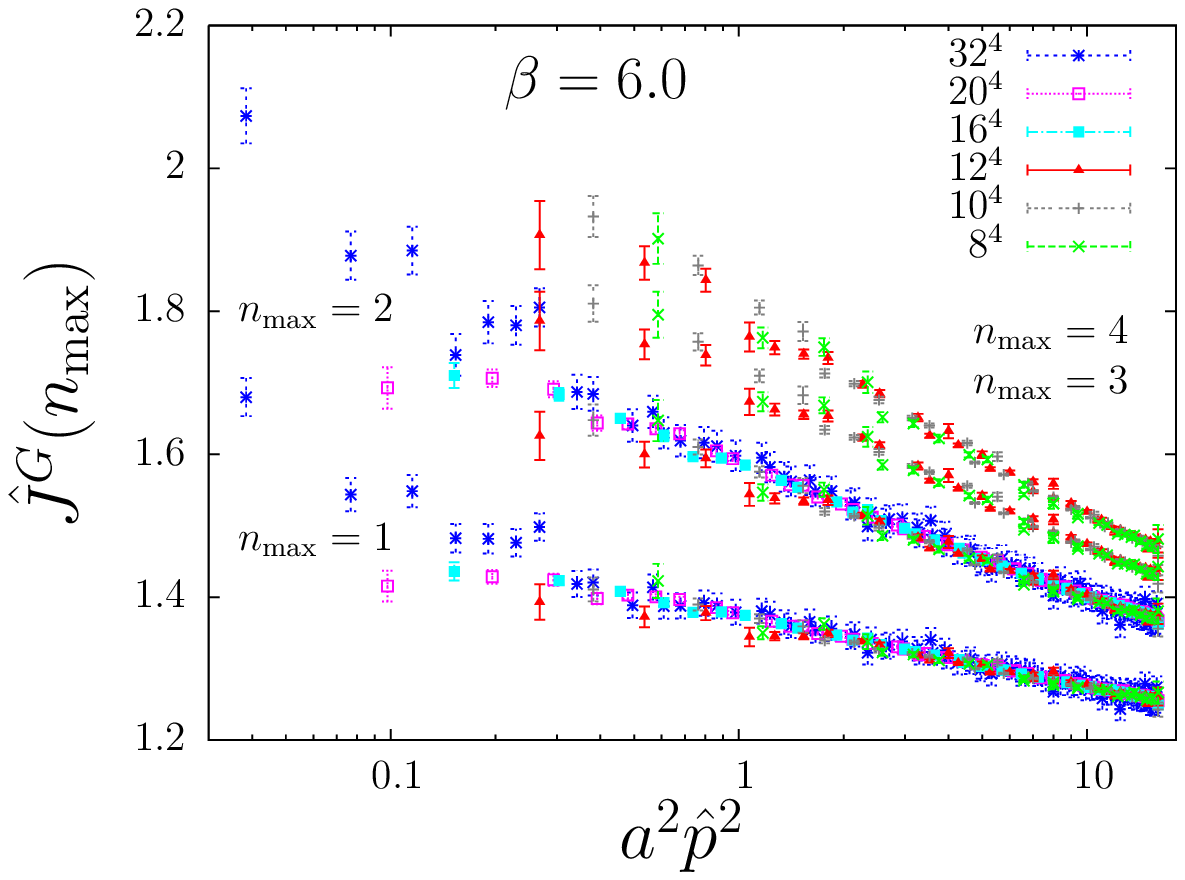}
       &
       \includegraphics[scale=0.64,clip=true]{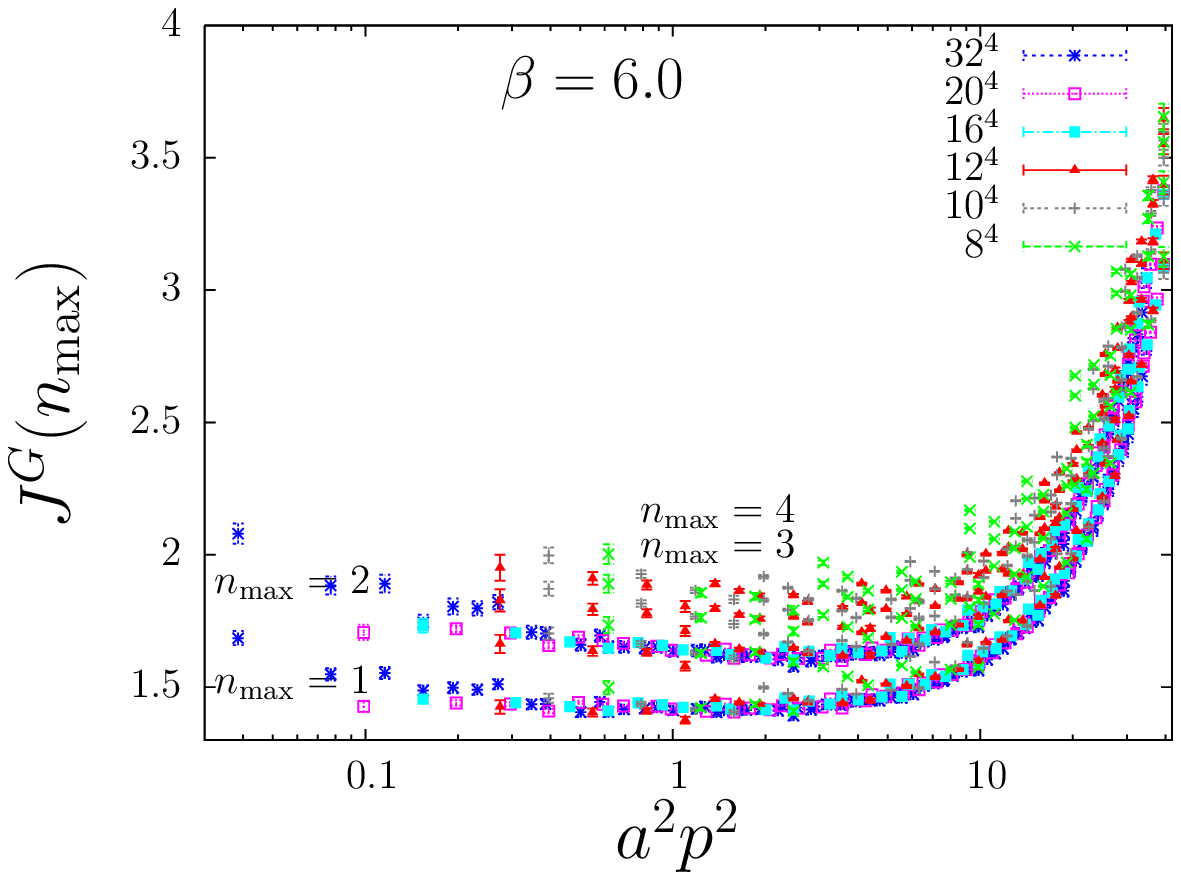}
    \end{tabular}
  \end{center}
    \caption{The cumulatively summed perturbative dressing function $\hat J^G(n_{\max})$ (left)
              and  $J^G(n_{\max})$ (right) up to four loops
              (two loops) using $\beta=6$ at $N=8,10,12$ ($16,20,32$).}
     \label{fig:summed-loops6}
\end{figure}

\begin{figure}[!htb]
  \begin{center}
    \begin{tabular}{cc}
       \includegraphics[scale=0.64,clip=true]{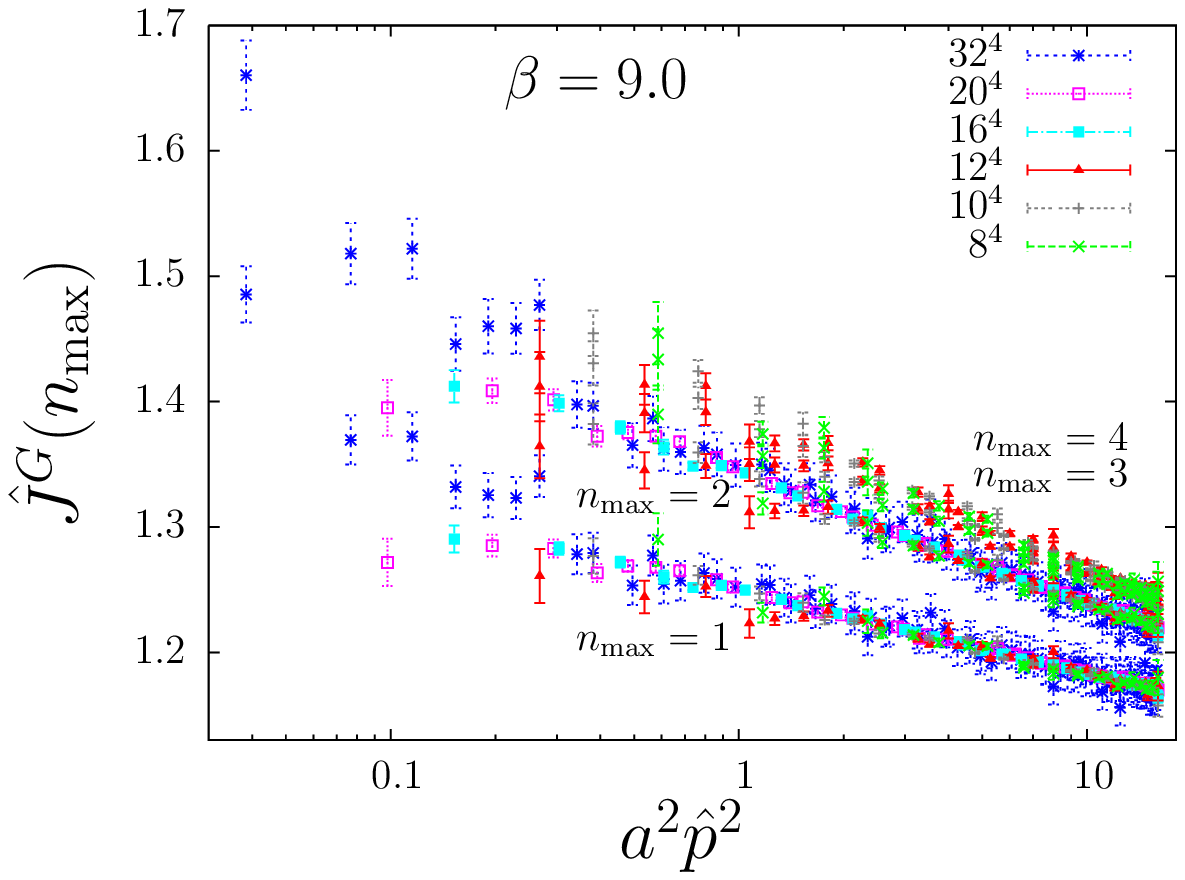}
       &
       \includegraphics[scale=0.64,clip=true]{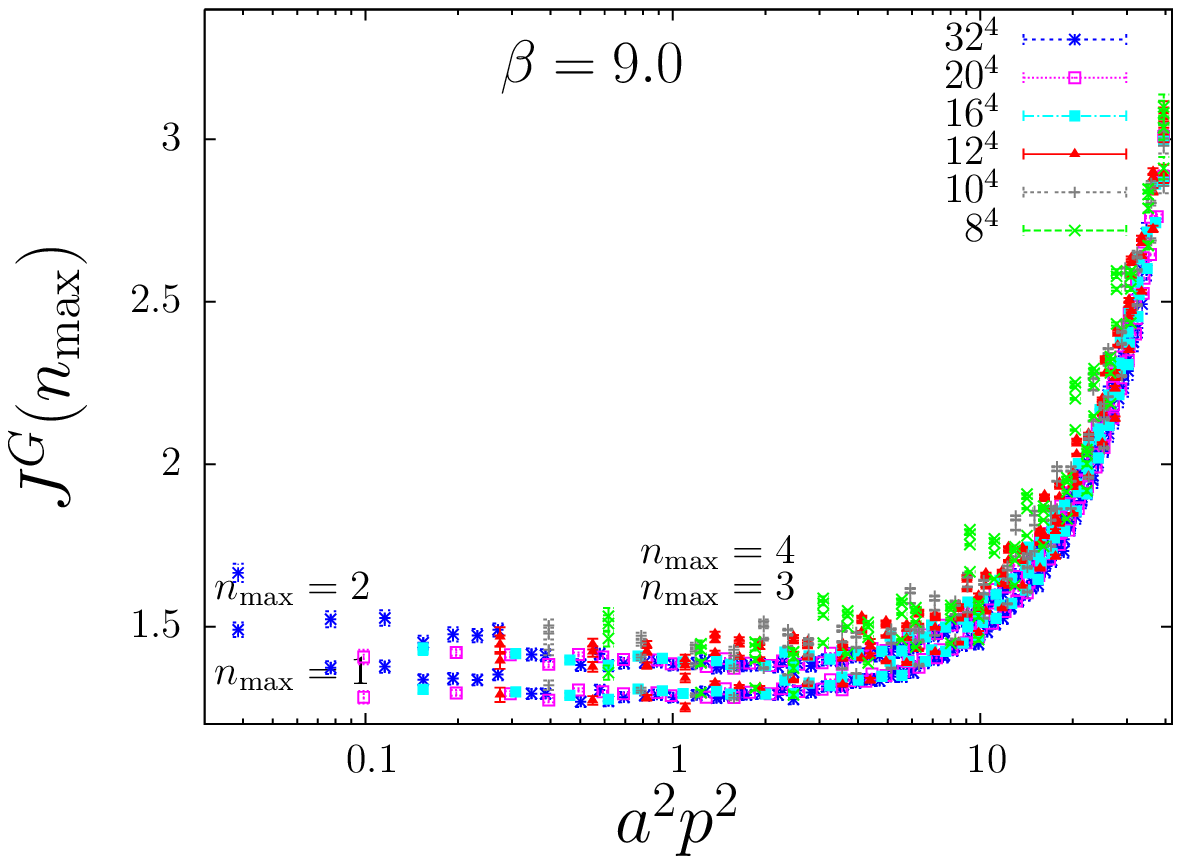}
    \end{tabular}
  \end{center}
    \caption{Same as in Fig.~\ref{fig:summed-loops6} at $\beta=9$.}
            \label{fig:summed-loops9}
\end{figure}

Restricting ourselves to four-momentum tuples near the diagonal we obtain the following results 
summarized in Figs.~\ref{fig:summed-loops6} and~\ref{fig:summed-loops9}.
It is interesting to notice that all loop contributions are of the same sign. 
Therefore, the summed dressing function most probably represents a lower bound 
on the complete perturbative dressing function.
We have to stress that our perturbative results at large $a^2\hat p^2$ or $(a p)^2$, 
being at the edges of the Brillouin zones, 
have nothing to do with the continuum limit in the ultraviolet
but could eventually describe the Monte Carlo data in  that large lattice momentum region.

\subsection{Summation using boosted perturbation theory and comparison with Monte Carlo results}

It is well-known that the bare lattice coupling $g^2=6/\beta$ is a bad expansion
parameter. Due to lattice artifacts like non-vanishing tadpole graphs, 
the lowest order perturbative coefficients are typically very large.
Therefore, it has been proposed~\cite{Lepage:1992xa} to use a boosted coupling $g_b^2$ instead
of the bare $g^2$ to improve the relative convergence behavior.
We use here a variant of boosting
where $g_b^2$  is computed via the perturbative plaquette $P_{\rm pert}(g^2)$ calculated in NSPT
(instead of using plaquette measurements from Monte Carlo simulations):
\begin{equation}
  g_b^2 = \frac{g^2}{P_{\rm pert}(g^2)}\,.
  \label{gboost}
\end{equation}

Let us expand the perturbative dressing function $\hat J^{G}(n_{\max})$ in $g^2$
and write
\begin{equation}
  \hat J^{G}(n_{\max})=  \sum_{n=0}^{n_{\max}} g^{2n} \, \hat j^{G,(n)}\,.
  \label{JGnaive}
\end{equation}
The analogous expansion in the boosted coupling $g_b^2$ has the form
\begin{equation}
  \hat J_b^{G}(n_{\max})=  \sum_{n=0}^{n_{\max}} g_b^{2n} \, \hat j_b^{G,(n)}\,.
  \label{JGBoost}
\end{equation}
Inserting 
\begin{equation}
P_{\rm pert}(g^2,n_{\max})=  1+ \sum_{n=1}^{n_{\max}} g^{2n} \, c_n
\label{Ppert}
\end{equation}
into (\ref{gboost}) we compute the boosted expansion coefficients $\hat j_b^{G,(n)}$ as functions 
of the naive coefficients $\hat j^{G,(n)}$ and $c_n$.
For the orders under consideration we thus obtain
\begin {eqnarray}
  \label{eq:reordering}
  \hat j_b^{G,(0)} &=& \hat j^{G,(0)} \,,
  \nonumber
  \\
  \hat j_b^{G,(1)} &=& \hat j^{G,(1)} \,,
  \nonumber
  \\
  \hat j_b^{G,(2)} &=& \hat j^{G,(2)} +  c_1 \, \hat j^{G,(1)} \,,
  \\
  \hat j_b^{G,(3)} &=& \hat j^{G,(3)} + 2\,c_1 \, \hat j^{G,(2)}+ ( c_1^2 + c_2)\, \hat j^{G,(1)}  \,,
  \nonumber
  \\
  \hat j_b^{G,(4)} &=& \hat j^{G,(4)} + 3\,c_1 \, \hat j^{G,(3)}  + ( 3\, c_1^2 + 2\, c_2) \,  \hat j^{G,(2)}  
  +  ( c_1^3 + 3\,  c_1 \,c_2+  \, c_3)\, \hat j^{G,(1)} 
  \,.
  \nonumber
\end {eqnarray}

Of course, for the infinite series both expansions should coincide $\hat J^{G}(\infty)=\hat J_b^{G}(\infty)$ --
but for every finite, truncated order $(n_{\max} < \infty)$ the perturbative series differ. 
Since the plaquette is less than one, it is clear from (\ref{gboost}) that $g_b^2 > g^2$.
However, the boosted expansion coefficients become significantly smaller than their naive 
counterparts, $|\hat j_b^{G,(n)}| \ll |\hat j^{G,(n)}|$.
The combination of both effects results in an improved convergence behavior of the series (\ref{JGBoost}).

An example of such a reordering using (\ref{eq:reordering}) is given in Table~\ref{tab:reorder} 
\begin{table}[!htb]
  \begin{center}
     \begin{tabular}{|c|c|c|}
       \hline 
                       &                   &                   \\ [-1.5ex]
       loop order $n$ &  $\hat j^{G,(n)}$ & $\hat j_b^{G,(n)}$\\ [0.5ex]
       \hline
                  0    &      0.9804(85)   &   0.9804(85)   \\  
                  1    &      0.3638(44)   &   0.3638(44)   \\
                  2    &      0.1995(31)   &   0.07762(165) \\
                  3    &      0.1296(24)   &   0.02445(61)  \\
                  4    &      0.09070(193) &   0.007847(323)\\
       \hline
     \end{tabular}
  \end{center}
  \caption{Expansion coefficients  $j^{G,(n)}$ and  $j_b^{G,(n)} $ defined in  
            (\ref{JGnaive}) and (\ref{JGBoost}), respectively, 
            for $\hat J^{G}(n_{\max}=4)$ assigned to the
            momentum tuple $(1,1,1,1)$ and the lattice size $N=12$.} 
  \label{tab:reorder}
\end{table}
for $\hat J^{G}(n_{\max}=4)$ assigned to the momentum tuple
$(1,1,1,1)$  at lattice size $N=12$. For the expansion coefficients of the plaquette in NSPT
we use the measured $c_n$~\cite{Ilgenfritz:2009ck} 
\begin{equation}
  c_1= -0.334998\,,  \quad
  c_2= -0.0337441 \, \quad
  c_3= -0.0137452 \, \quad
  c_4= -0.00729851
\end{equation}
for the same lattice size $N=12$ (neglecting their tiny errors).
The two-, three- and four-loop expansion coefficients given in Table~\ref{tab:reorder} 
are significantly smaller using boosted perturbation theory.

\begin{figure}[!htb]
  \begin{center}
    \begin{tabular}{cc}
       \includegraphics[scale=0.64,clip=true]{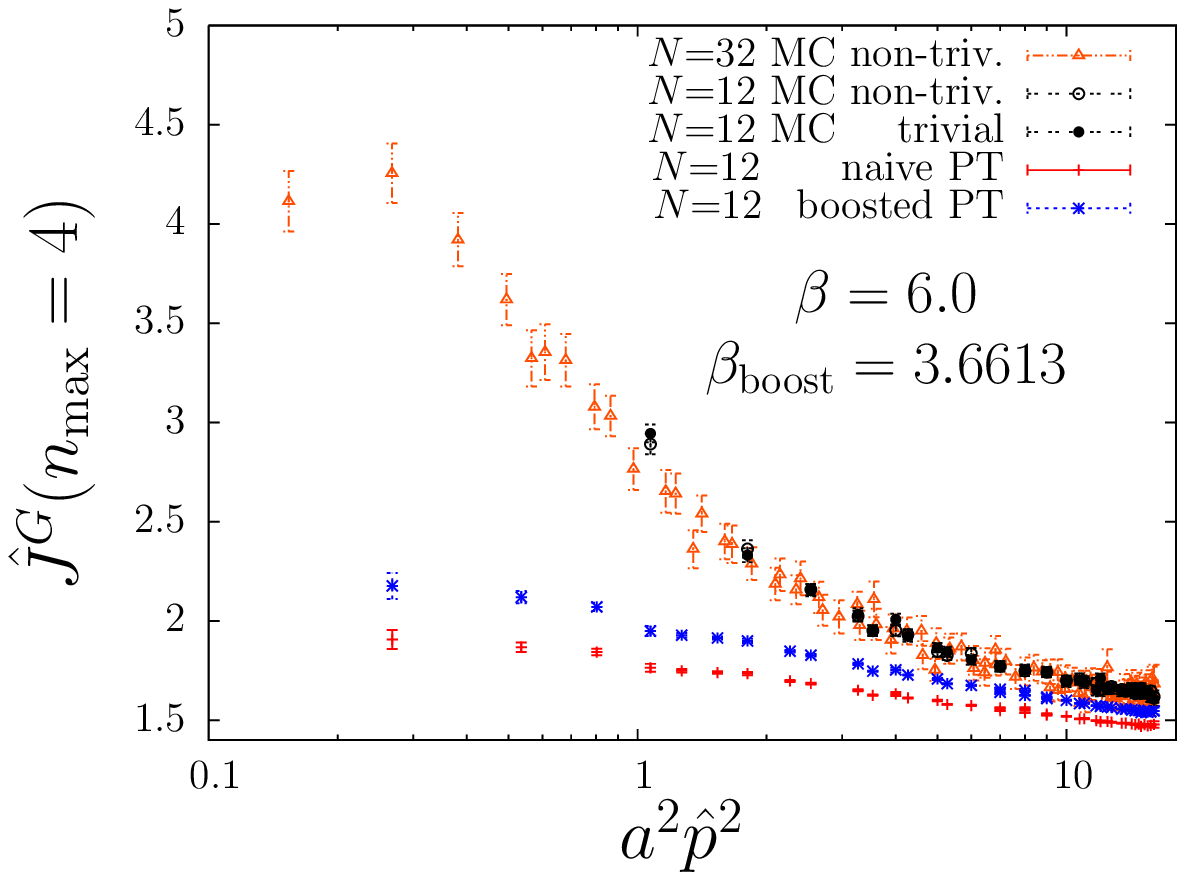}
       &
       \includegraphics[scale=0.64,clip=true]{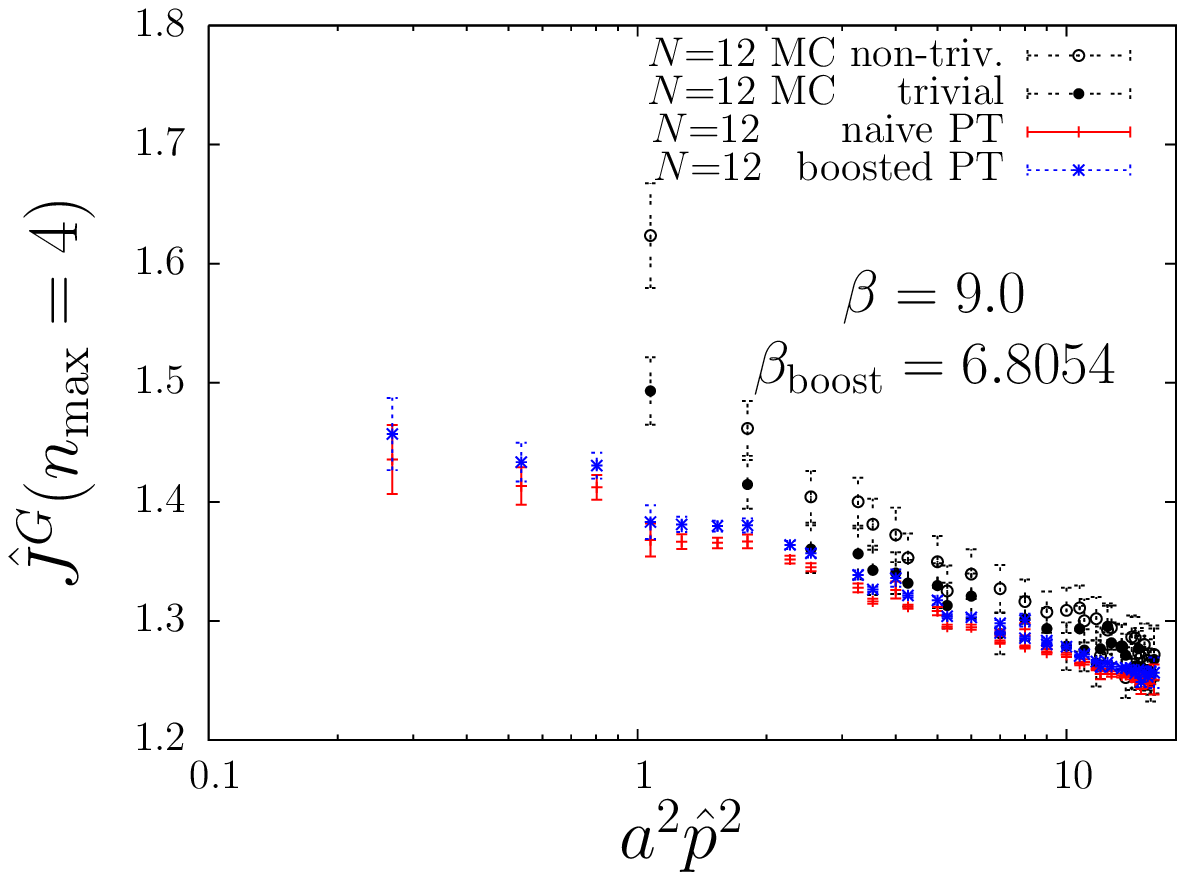}
    \end{tabular}
  \end{center}
  \caption{Comparison of naive and boosted perturbation theory for the gluon dressing 
           function at $N=12$ with Monte Carlo data including real and complex Polyakov loop
           sectors (non-trivial) and restricted to the real (trivial) Polyakov sector from~\cite{Menz} 
           at two different $\beta$ values.
           To show the non-perturbative low momentum behavior, also $N=32$ Monte Carlo data are included
           at $\beta=6$.}
   \label{fig:summed-boostedloops}
\end{figure}

In Figs.~\ref{fig:summed-boostedloops} 
we show the effect of boosting for the dressing function
$\hat J^{G}(n_{\max}=4)$ at two different $\beta$ values
and compare it with Monte Carlo results of the Berlin Humboldt-University group~\cite{Menz}
which has used the same gauge field definition (I-6) as implied in NSPT.
The boosted result $\hat J_b^{G}(n_{\max}=4)$  is shifted towards the Monte Carlo results.
Note that at the larger $\beta$ value the Monte Carlo results significantly depend 
on whether the measurements are restricted to the real (trivial) Polyakov sector.
We observe a reasonable agreement of the NSPT summed dressing function  
at the largest  $(a^2 \hat p^2)$ with the Monte Carlo data obtained 
in the trivial Polyakov sector for all four directions 
(if necessary, reached by $Z(3)$ flips). For details, see~\cite{Menz}.

In contrast to Monte Carlo results~\cite{Sternbeck:2005tk}, we do not see even in higher-loop 
lattice perturbation theory any sign for 
a suppression for $\hat J$ in the infrared direction (at small  $a^2\hat p^2$), 
such that this principal difference must be related 
to non-perturbative effects responsible for the expected confinement behavior 
of the gluon propagator.
One of interesting questions might be to what extent the difference could be 
effected by some phenomenological constants. 
The gluon condensate might be a prominent example.  

\section{Fit results for LPT in the infinite volume limit}
\label{sec:fitting}

Analogously to paper I we extract the finite constants $J^G_{i,0}$ for loop order $i$
in the perturbative expansion of the
lattice gluon dressing function (\ref{Zgluonbeta3loop}) in Landau gauge.
We use the same strategy as outlined there.

First we subtract all logarithmic pieces (supposed to be universal and known)
from the gluon dressing function for each momentum tuple and for all lattice sizes.
Next we select a range in
$(ap)^2= \sum (ap_\mu)^2$, $a p_\mu= k_\mu (2 \pi/N)$ with
$(ap)_{\min}^2 < (ap)^2 < (ap)_{\max}^2$.
Within that range we identify a set of four-momentum tuples $(k_1,k_2,k_2,k_4)$
which is common to all chosen lattice sizes.
The data in that set are assumed to have the same $pL = p \, (aN)$ effects 
for each given momentum tuple.
Since finite-volume effects decrease with increasing momentum squared,
we  choose as reference fitting point -- for an assumed behavior at
$N=\infty$ -- an additional data point at $(ap)^2 \approx (ap)^2_{\max}$ from the largest
lattice size at our disposal.
Then we  perform a non-linear fit using all data points 
from different lattice sizes $N$ in that set plus the reference point
correcting for finite size (no functional form guessed)
and assuming a specific functional behavior for the $H(4)$ dependence.
This functional form analogously to (I-62) is a hypercubic-invariant Taylor series 
(with $(ap)^n=\sum_\mu (a p_\mu)^n$)
\begin{equation}
  J^G_{i,0}(ap) =  J^G_{i,0}
  + c_{i,1} \, (ap)^2
  + c_{i,2} \, \frac{(ap)^4}{(ap)^2}
  + c_{i,3} \, (ap)^4
  + c_{i,4} \, \left((ap)^2\right)^2
  + c_{i,5} \, \frac{(ap)^6}{(ap)^2}
  + \cdots \,.
  \label{eq:H4TaylorG}
\end{equation}
Here $J^G_{i,0}(ap)$ denotes the part of the dressing function 
which does not depend on $pL = p \, (aN)$ and logarithmic effects.
Finally we vary the momentum squared window and find an optimal $\chi^2$
region which allows us to find the ``best'' $J^G_{i,0}$.

This time we use our NSPT data  -- extrapolated to zero Langevin step -- 
from lattice sizes $\{N = 8, 10, 12, 16, 20, 32\}$ for the one-- and two-loop case.
We take  into account tuples with $|k_i-k_j| \le \Delta k_{\max}$ similar to (I-58).
In addition to the ghost propagator case some 
more tuples are allowed,
{\em e.g.}  $\Delta k_{\max}=2$:  (2,0,0,0), (3,3,1,1), \dots

Since we have now a larger lattice volume available 
when setting the scale as explained in I,
the possible optimal fitting window in the momentum squared $(ap)^2$ becomes larger 
and the number of four-momentum tuples fitted together from all lattices is increased.
To get a rough estimate of the three-loop finite constant, we also use the available
NSPT measurements from smaller lattice sizes $\{N = 6, 8, 10, 12\}$.

\begin{figure}[!htb]
  \begin{center}
    \begin{tabular}{cc}
       \includegraphics[scale=0.64,clip=true]{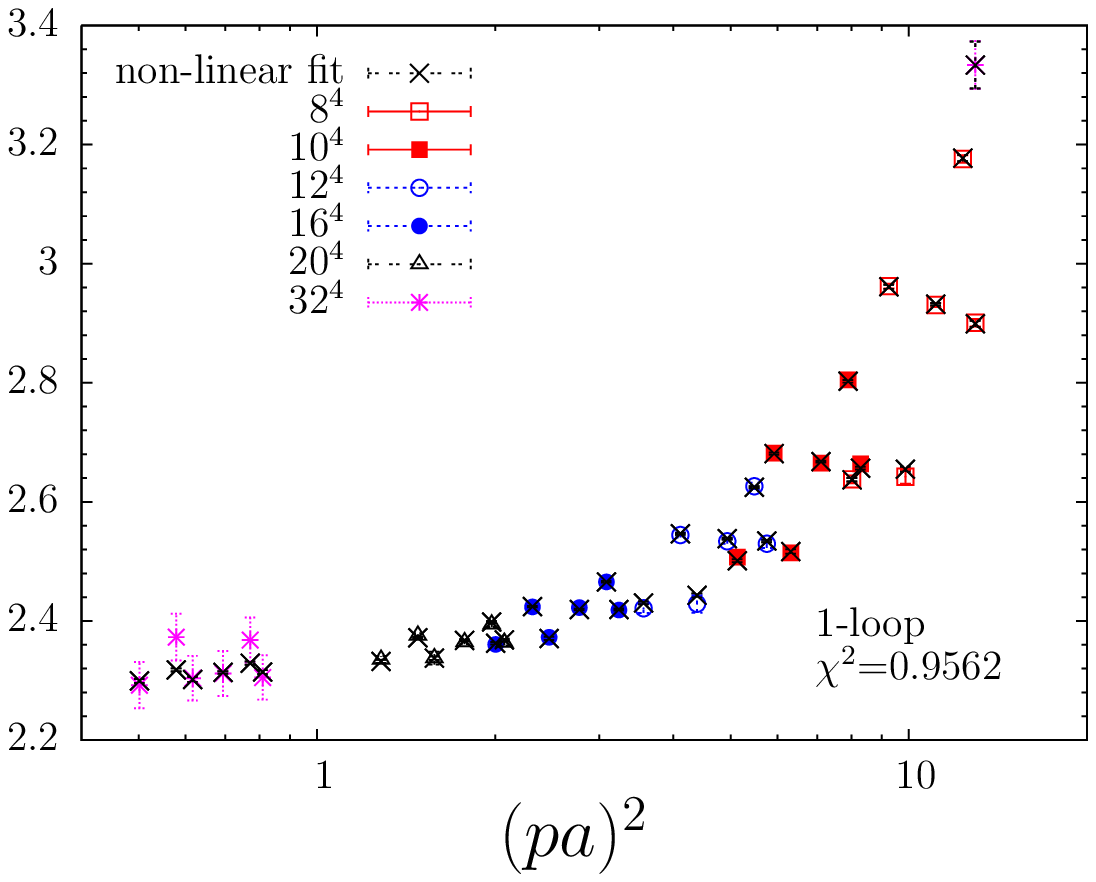}
&
       \includegraphics[scale=0.64,clip=true]{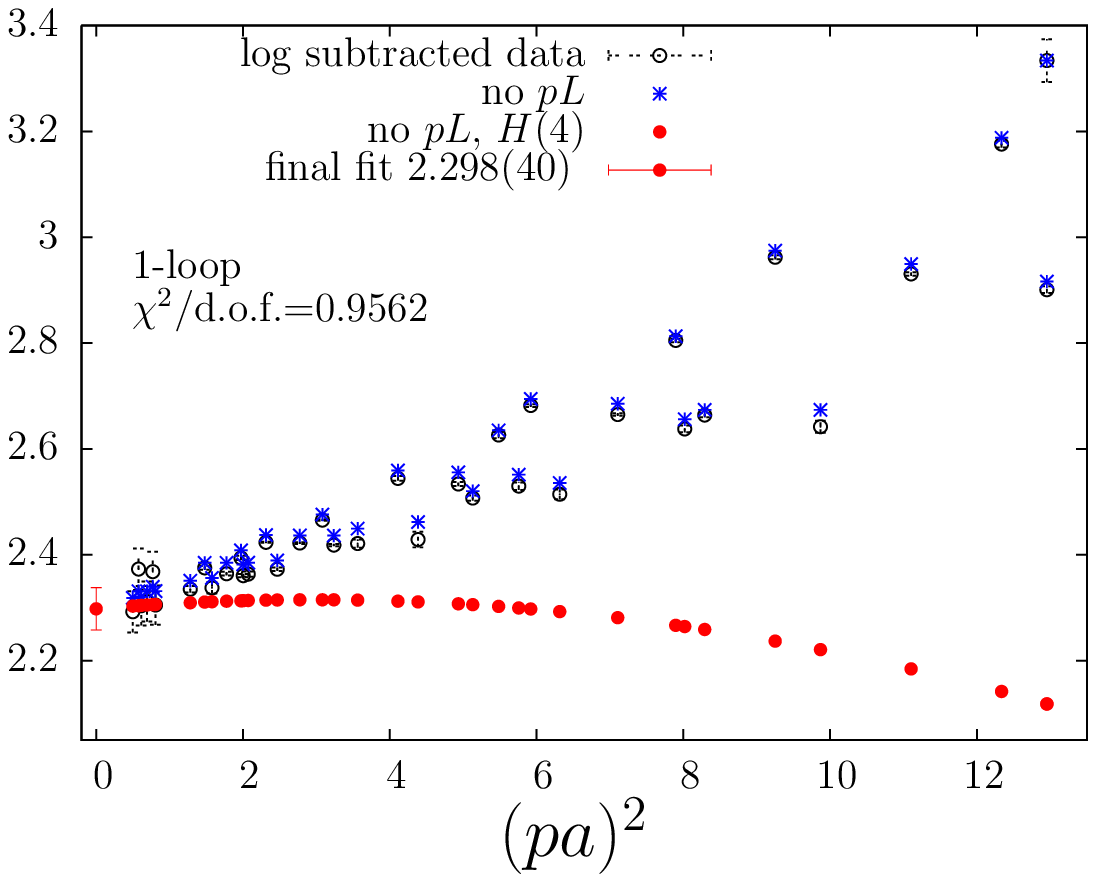}
    \end{tabular}
  \end{center}
  \caption{Fitting of the one-loop non-logarithmic coefficient. 
           Left:  Data points after subtracting logarithms for various lattice sizes 
                  compared to points using the non-linear fit of type (I-66) with 
                  (\ref{eq:H4TaylorG}) analogous to (I-62).
           Right: Stars denote data after correction for finite-volume effect,
                  the finite volume effect, {\it i.e.} they represent $J^G_{i,0}(ap)$; 
                  full circles are fit points after correcting both finite volume
                  and some hypercubic effects with the exception of those proportional 
                  to the coefficients $c_{i,1}$ and $c_{i,4}$ in (\ref{eq:H4TaylorG}).}
  \label{fig:gpLatWork1}
\end{figure}

\begin{figure}[!htb]
  \begin{center}
    \begin{tabular}{cc}
       \includegraphics[scale=0.64,clip=true]{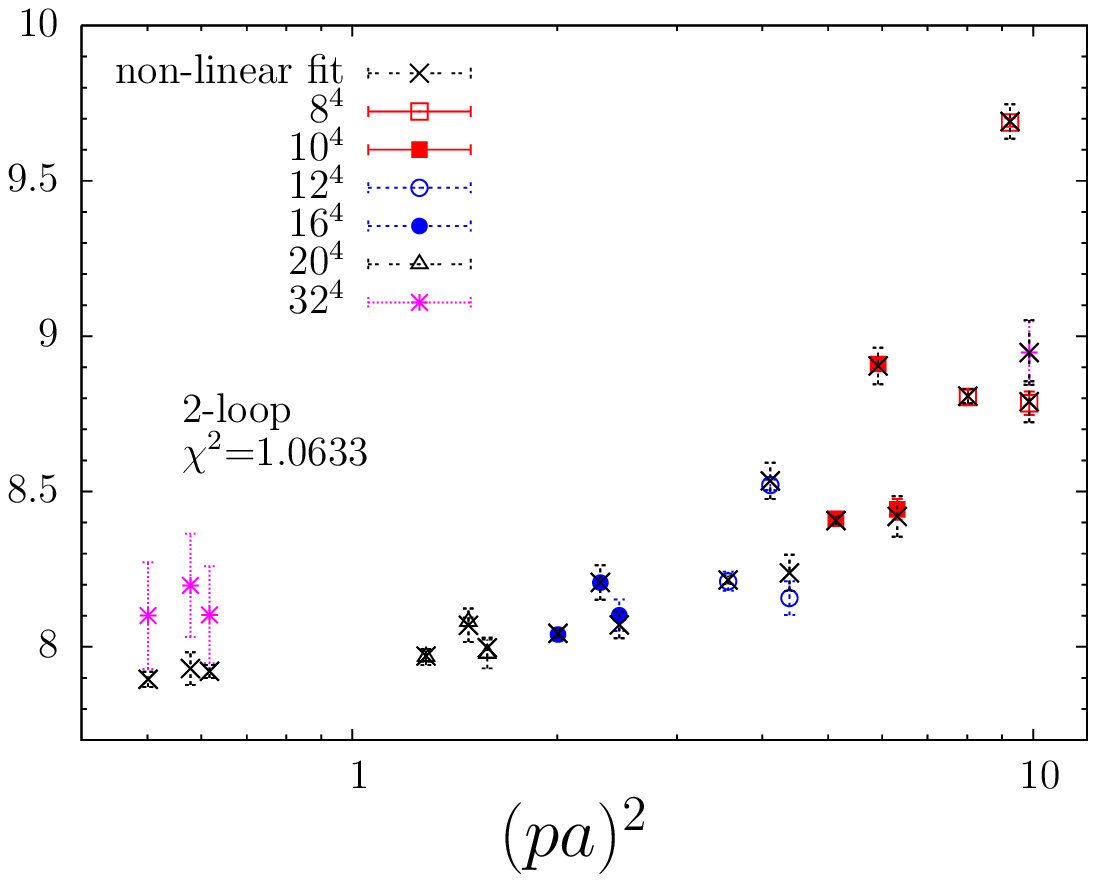}
&
       \includegraphics[scale=0.64,clip=true]{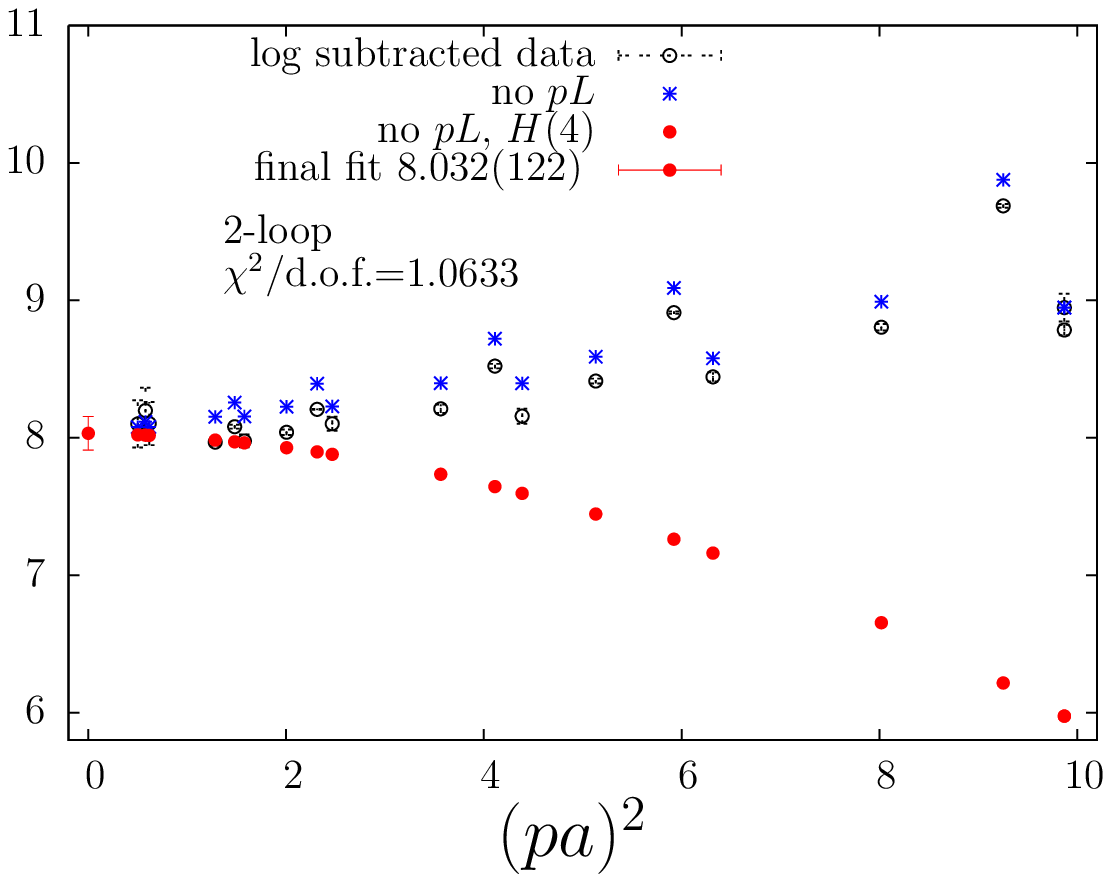}
    \end{tabular}
  \end{center}
  \caption{Same as in Fig.~\ref{fig:gpLatWork1} for the two-loop non-logarithmic coefficient.}
  \label{fig:gpLatWork2}
\end{figure}

\begin{figure}[!htb]
  \begin{center}
    \begin{tabular}{cc}
       \includegraphics[scale=0.64,clip=true]{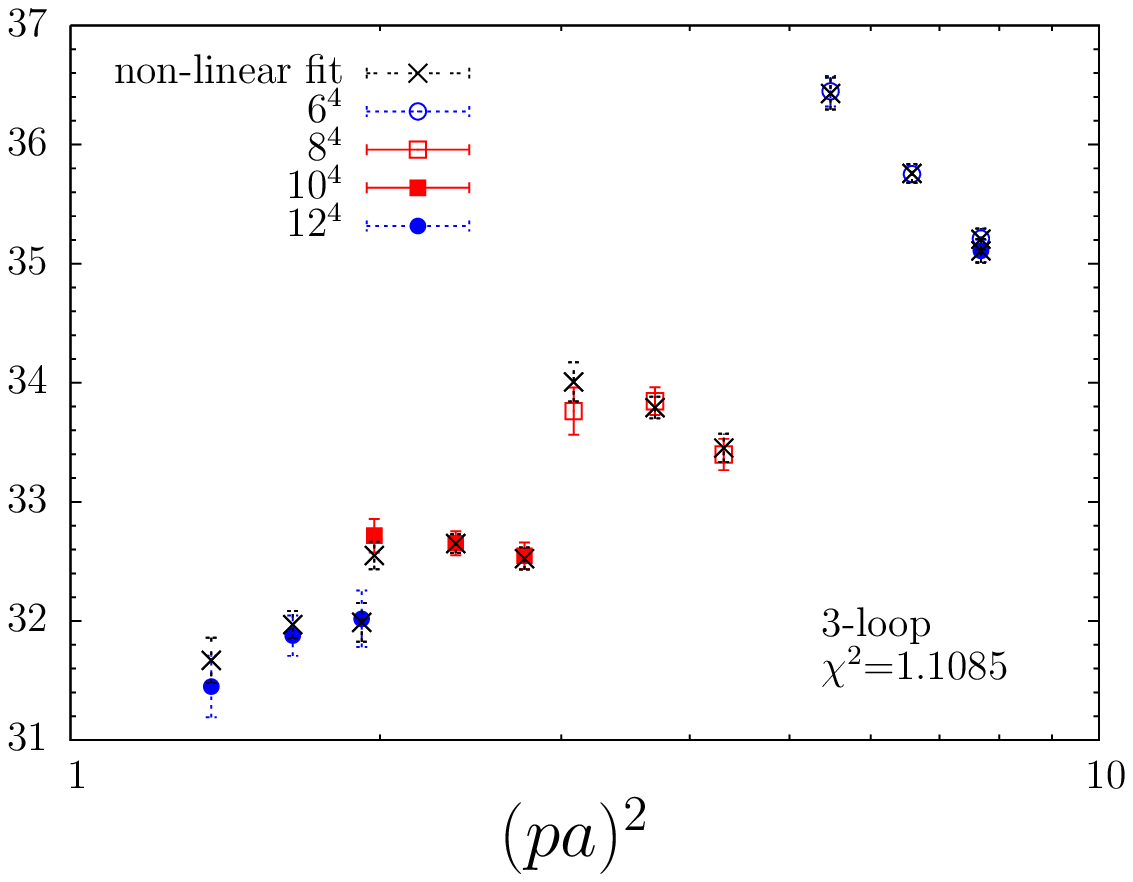}
&
       \includegraphics[scale=0.64,clip=true]{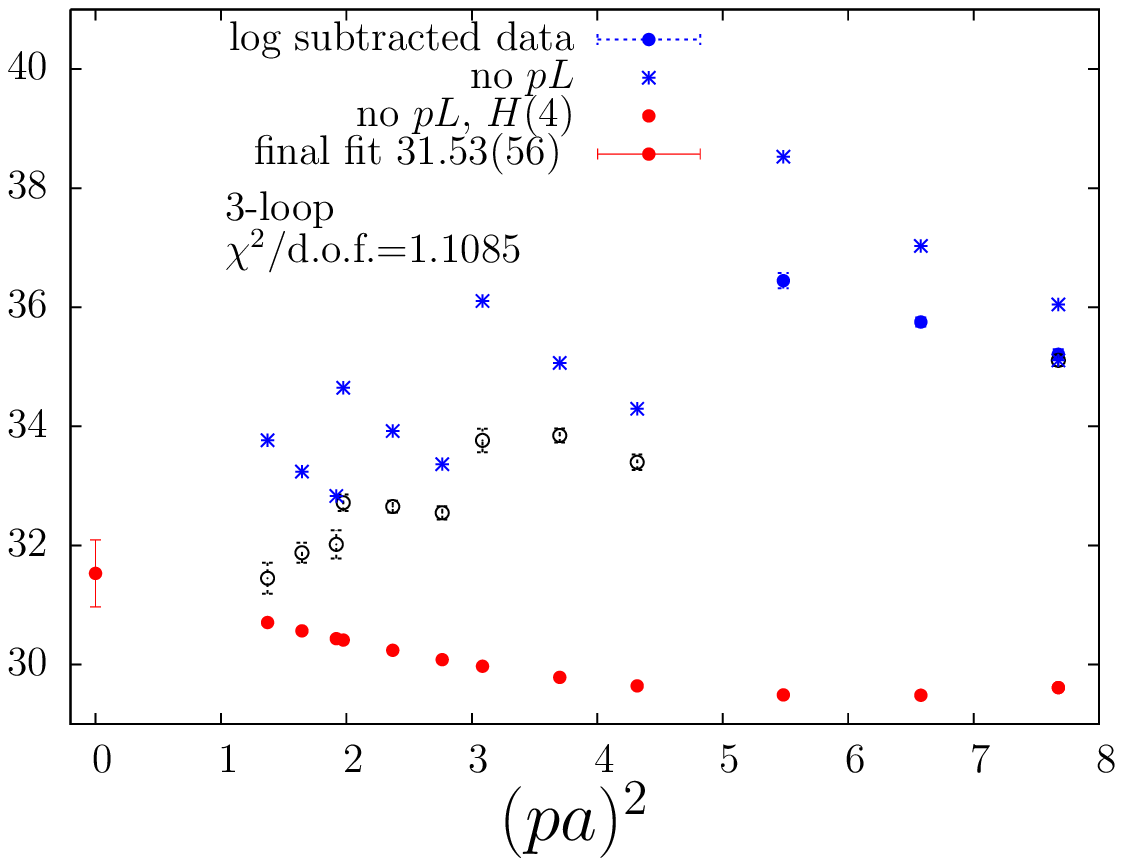}
    \end{tabular}
  \end{center}
  \caption{Same as in Fig.~\ref{fig:gpLatWork1} for the three-loop non-logarithmic coefficient.}
  \label{fig:gpLatWork3}
\end{figure}

As examples of how the fitting  works
we present in Figs.~\ref{fig:gpLatWork1}, \ref{fig:gpLatWork2}  and \ref{fig:gpLatWork3} 
the handling of one-loop, two-loop and three-loop gluon dressing functions at a particular $\chi^2$ value
switching off the estimated lattice artifacts in two steps: first the $(pL)$-effect then the $H(4)$
dependence.

\begin{figure}[!htb]
  \begin{center}
       \includegraphics[scale=0.63,clip=true]{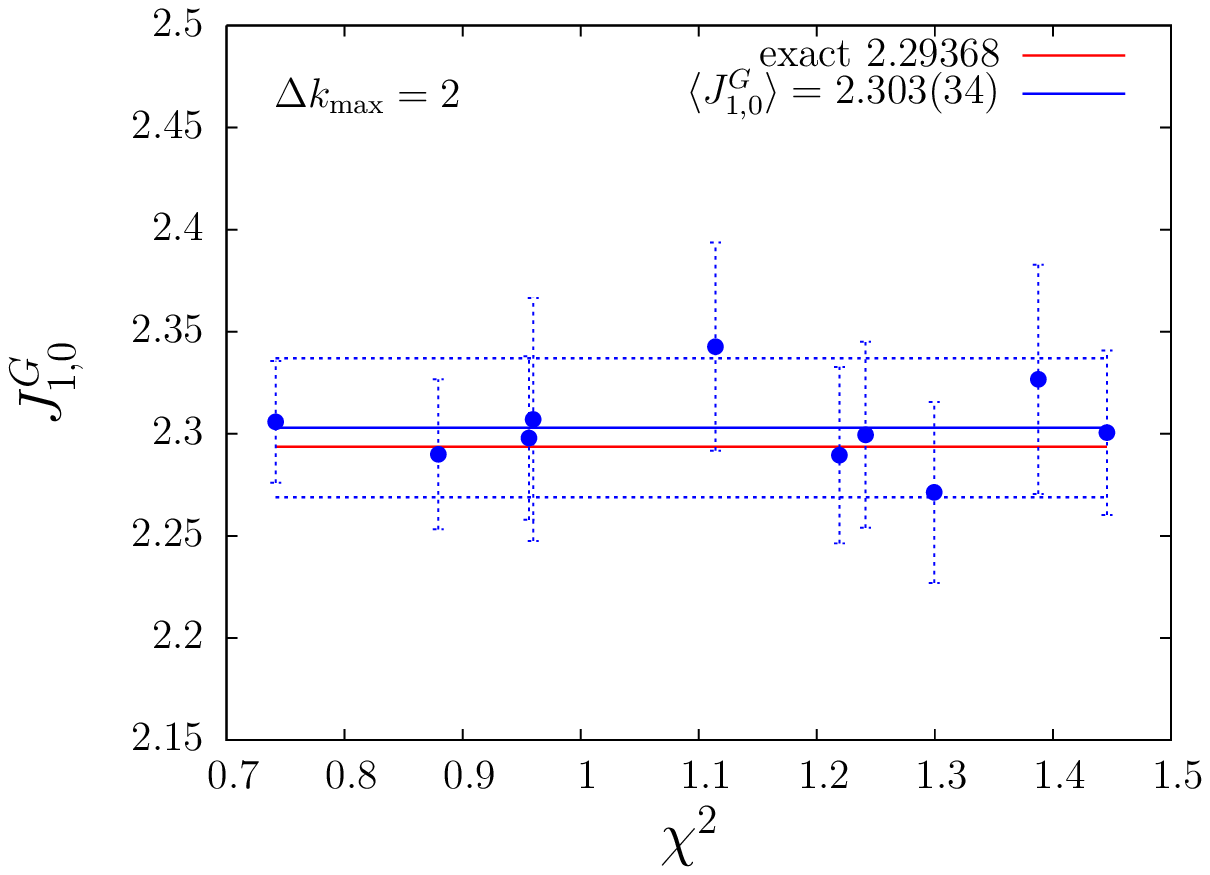}
  \end{center}
  \caption{One-loop results $J^G_{1,0}$ allowing $\Delta k_{\max}=2$ for the smallest available
           $\chi^2$ together with the exact value 2.29368 (see (\ref{JG1loop})).}
  \label{fig:OneLoopGFit}
\end{figure}

\begin{figure}[!htb]
  \begin{center}
    \begin{tabular}{cc}
       \includegraphics[scale=0.63,clip=true]{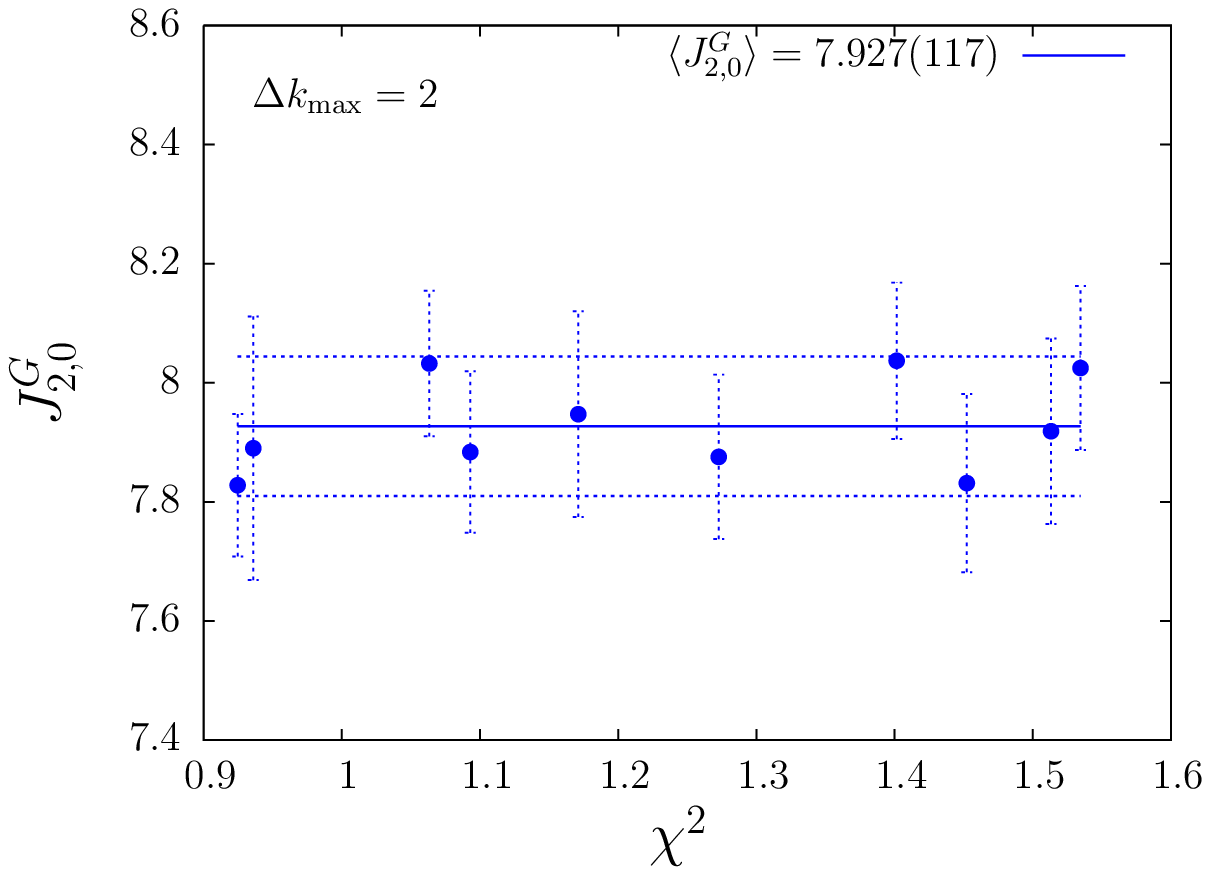}
&
       \includegraphics[scale=0.63,clip=true]{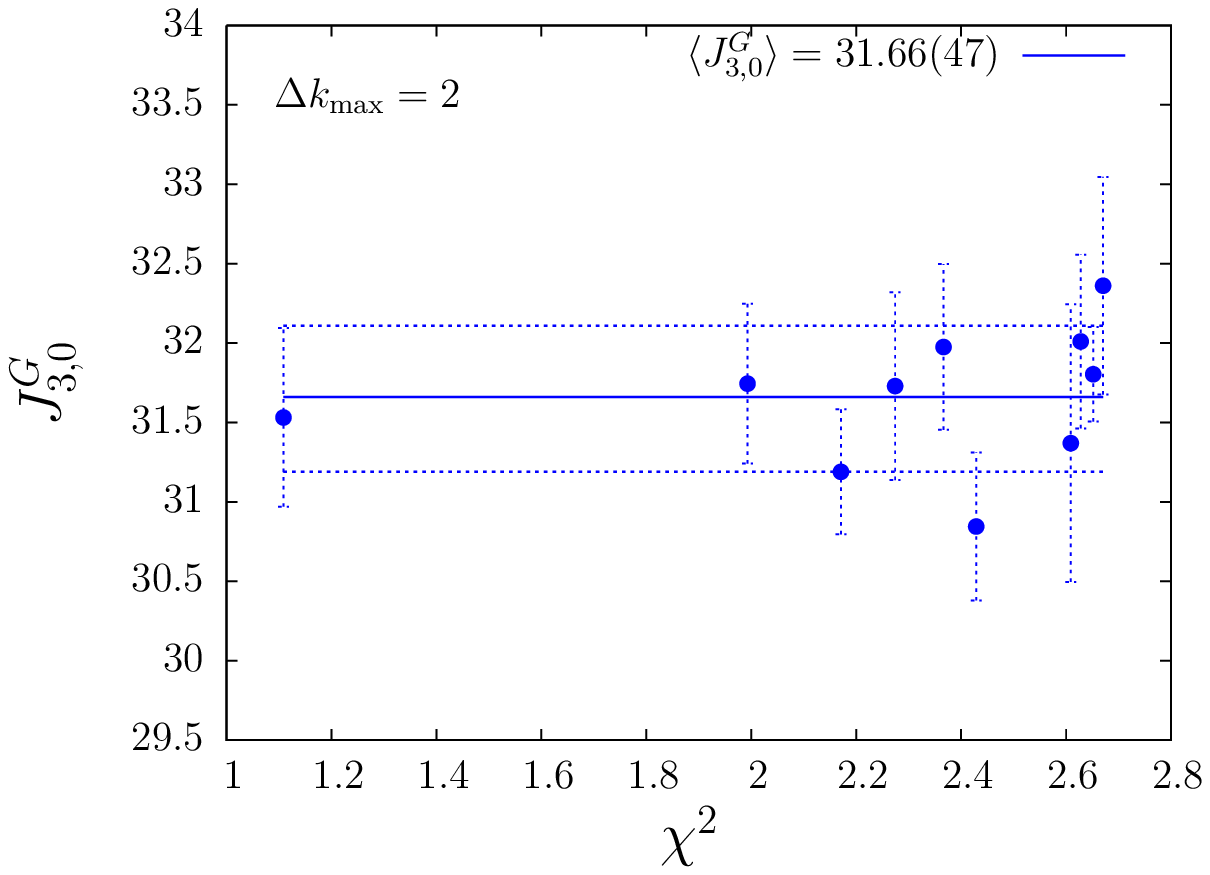}
    \end{tabular}
  \end{center}
  \caption{Similar to Fig.~\ref{fig:OneLoopGFit} for the analytically unknown 
           two-loop and three-loop non-logarithmic coefficients $J^G_{2,0}$  and $J^G_{3,0}$.}
  \label{fig:TwoThreeLoopGFit}
\end{figure}

The results for the one-loop constant for ten best $\chi^2$ -values using  $\Delta k_{\max}=2$ are shown in 
Fig.~\ref{fig:OneLoopGFit}
which agree within errors. 
The corresponding best two- and three-loop constants for $\Delta k_{\max}=2$ are given in 
Fig.~\ref{fig:TwoThreeLoopGFit}.
Note that for individual  $\chi^2$ values
the momentum squared windows differ, depending on the number of
contributing four-momentum tuples which enter into the non-linear fits.

Table~\ref{tab:fitsummary} contains a summary of the non-logarithmic constants $J^G_{i,0}$
we have found using three different selection criteria $\Delta k_{\max}=1, 2, 3$ for 
the four-momentum tuples.
\begin{table}[!htb]
  \begin{center}
     \begin{tabular}{|c|ccc|}
        \hline
                                   &               &$\Delta k_{\max}$&           \\[0.5ex]
        \cline{2-4}
        &&& \\ [-2.0ex]
                                   &         1     &        2        &     3     \\ [0.5ex]
        \hline
        &&& \\ [-2.0ex]
        $\langle J^G_{1,0}\rangle$ &  2.318(56)    &  2.303(34)      & 2.292(69) \\  
        &&& \\ [-2.0ex]
        $\langle J^G_{2,0}\rangle$ &  7.939(123)   &  7.927(117)     & 7.897(112)\\ 
        &&& \\ [-2.0ex]
        $\langle J^G_{3,0}\rangle$ &  31.61(33)    &   31.66(47)     & 31.56(46) \\  
        \hline
     \end{tabular}
  \end{center}
  \caption{Non-logarithmic constants $\langle J^G_{i,0}\rangle$ of the gluon dressing function
           averaged over the best ten $\chi^2$ values of the non-linear fits 
           using different selection criteria $\Delta k_{\max}$ for the four-momentum tuples.}
  \label{tab:fitsummary}
\end{table}
The errors are estimated 
by equally weighting the mean deviations squared of both the individual fits and the sum of the
``best'' ten $\chi^2$ values. 
The results from the different criteria coincide within errors.
We have obtained a very good agreement at the level of 0.5 percent
with the expected exact one-loop result  $J^G_{1,0}$  given in~(\ref{JG1loop}).

Comparing the accuracy reached here with that of the ghost propagator, 
let us mention again that in the gluon propagator case already the
tree level is calculated from quantum fluctuations of the gauge fields.
So the one-loop accuracy in the gluon dressing function case should be 
fairly compared to the two-loop accuracy of the ghost dressing function.

To estimate the influence of the missing larger volumes for the three-loop constant, 
let us compare the two-loop constants obtained in the lattice volume sets ($N=8,\dots,32$) 
(one- and two-loop) and ($N=6,\dots,12$) (three-loop).
The results are collected in Table ~\ref{tab:fitcomp}.
\begin{table}[!htb]
  \begin{center}
     \begin{tabular}{|c|c|c|}
        \hline
        \\[-2.0ex]
        $\Delta k_{\max}$ &$\langle J^G_{2,0}\rangle$ &$\langle J^G_{2,0}\rangle$
        \\[0.5ex]
                          &   $N=8,\dots,32$ &  $N=6,\dots,12$
        \\[0.5ex]
        \hline
         1                &    7.939(123)    &                 7.919(79)   \\ 
         2                &    7.927(117)    &                 7.876(106)  \\ 
         3                &    7.897(112)    &                 7.888(115)  \\ 
        \hline
     \end{tabular}
  \end{center}
  \caption{Comparison of the two-loop constant  $\langle J^G_{2,0}\rangle$ using 
           two different lattice volume sets.}
  \label{tab:fitcomp}
\end{table}
{}From the numbers given there we conclude that the missing data sets in the three-loop fit  of  
$\langle J^{G,(3,0)}\rangle$ do not entail a significant change.
There is a small tendency to somewhat larger numbers using larger volumes. 
This has to be taken into account as a systematic effect in our estimate of  
$J^G_{3,0}$.

Finally we decided to take the selection criterion $\Delta k_{\max}=2$ as the most suitable one
and present our numerical results for the unknown non-logarithmic constants in the gluon dressing function 
of infinite volume lattice perturbation theory in Landau gauge:
\begin{eqnarray}
  J^G_{2,0}&=&7.93(12)\,,
  \\
  J^G_{3,0}&=& 31.7(5) \,.
\end{eqnarray}
Collecting all results we can write (\ref{Zgluonbeta3loop}) in a numerical form 
(restricting to at most five digits after the decimal point)
\begin{eqnarray}
  &&J^{G,\rm 3-loop}(a,p, \beta) = 1 + \frac{1}{\beta}\,\Bigl( -0.24697\, \log (ap)^2 + 2.29368 \Bigr)
  + \nonumber \\
  & &\hspace{-1cm} 
  + \frac{1}{\beta^2}\,\Bigl(0.08211\, \left( \log (ap)^2 \right)^2 - 1.48445 \, 
  \log (ap)^2 + 7.93(12) \Bigr) +
  \label{J3GloopNum} \\
  & & \hspace{-1cm}
  + \frac{1}{\beta^3}\,\Bigl( -0.02964\, \left( \log (ap)^2 \right)^3 + 0.81689 \, 
  \left( \log (ap)^2 \right)^2 - 8.13(3) \, \log (ap)^2 + 31.7(5) \Bigr)\,.
  \nonumber
\end{eqnarray}
A transformation to the $RI'$ scheme can be performed using the relations given in Section~\ref{sec:standard-LPT}.

 \section{Summary}
\label{sec:summary}
In the present work we have applied NSPT to calculate the Landau gauge gluon propagator in     
lattice perturbation theory up to four loops.
The summed gluon dressing function is compared to recent Monte Carlo measurements of the 
Berlin Humboldt University group. Both (NSPT) perturbative and non-perturbative results 
are in terms of one and the same definition of the gauge fields, both in Landau gauge fixing 
and in measurements of the propagator. To improve the comparison, we have also summed our results 
in a boosted scheme showing better convergence properties.

The key goal of the lattice study of propagators is to reveal their genuinely non-perturbative 
content, which asks for disentangling perturbative and non-perturbative contributions. The 
commonly used procedure goes through the fit of the high momentum tail by 
continuum-like formulae (anomalous dimensions and logarithms can be taken from continuum computation). 
While this lets us gain intuition, it opens the way to further ambiguities, since irrelevant 
effects give substantial contributions to the perturbative tail. At large lattice momenta our 
calculations indicate that the perturbative dressing function constructed by means of
NSPT with more than four loops will match the Monte Carlo measurements, thus enabling a fair 
accounting of the perturbative tail. 
The strong difference which is left over in the intermediate and -- moreover -- the infrared momentum 
region should then be attributed to non-perturbative effects. Power corrections~\cite{Boucaud:2008gn}
and contributions from non-perturbative excitations~\cite{Langfeld:2001cz,Gattnar:2004bf} are serious 
candidates for the description of these (better disentangled) effects. 

The one-loop result for the perturbative gluon propagator of Lattice $SU(3)$ in covariant gauges 
(and in particular Landau) has been known for a long time. Using our strategy for a careful 
analysis of finite volume and finite lattice size effects we find good agreement with this result.  
In (\ref{J3GloopNum}) we have summarized our (original) two- and three-loop results.

\section*{Acknowledgements}
This work is supported by DFG under contract SCHI 422/8-1, DFG SFB/TR 55,
by I.N.F.N. under the research project MI11 and by the Research Executive Agency (REA) 
of the European Union under Grant Agreement number PITN-GA-2009-238353 (ITN STRONGnet).

\end{document}